\documentclass[11pt,english]{article}
\usepackage{amsfonts, amsmath, amssymb, verbatim, setspace, pdfsync, graphicx,comment}
\usepackage{color}
\usepackage{pdfsync}
\usepackage{leftidx}
\usepackage{tikz}
\definecolor{darkgreen}{rgb}{0,0.5,0}
\definecolor{darkblue}{rgb}{0,0,0.6}
\definecolor{purple}{rgb}{0.4,0.15,0.21}
\definecolor{black}{rgb}{.2,.2,.2}
\usepackage[colorlinks=true,citecolor=darkgreen,linkcolor=black,urlcolor=darkblue]{hyperref}

\DeclareMathOperator{\p}{\partial}

\newcommand{\be}{\begin{equation}}
\newcommand{\ee}{\end{equation}}
\newcommand\eea{\end{eqnarray}}
\newcommand\bea{\begin{eqnarray}}
\newcommand{\f}{\frac}

\begin{document}
\unitlength = 1mm
\

\begin{center}

{ \LARGE \textsc{Timelike BKL singularities and chaos in ADS/CFT}}
\\
\vspace*{1.7cm}
Edgar Shaghoulian$^{B,K}$ and Huajia Wang$^{K,L}$\\
\vspace*{0.6cm}

\vspace*{0.6cm}
{\it ${}^B$ Department of Physics, UCSB, Santa Barbara, CA 93106}\\
{\it ${}^K$ Stanford Institute of Theoretical Physics, Stanford University\\}
{\it ${}^L$ Department of Physics, University of Illinois at Urbana-Champaign}

\vspace*{0.6cm}


\end{center}
\vspace{.5cm}

\begin{abstract}

\noindent We study the nature of a family of curvature singularities which are precisely the timelike cousins of the spacelike singularities studied by Belinski, Khalatnikov, and Lifshitz (BKL). We show that the approach to the singularity can be modeled by a billiard ball problem on hyperbolic space, just as in the case of BKL. For pure gravity, generic chaotic behavior is retained in $(3+1)$ dimensions, and we provide evidence that it disappears in higher dimensions. We speculate that such singularities, if occurring in AdS/CFT and of the chaotic variety, may be interpreted as (transient) chaotic renormalization group flows which exhibit features reminiscent of chaotic duality cascades. 

\end{abstract}

\pagebreak
\setcounter{page}{1}
\pagestyle{plain}

\setcounter{tocdepth}{1}

\tableofcontents

\section{Introduction}\label{intro}
It is remarkable that the intellectual environment created by the Cold War could have a positive effect on the practice of science. Distinct approaches to problems were developed instead of singular techniques dominating the field. One of the beautiful pieces of theoretical physics to emerge from the isolated intellectual environment of the former Soviet Union is the analysis by Belinski, Khalatnikov, and Lifshitz (BKL) of the nature of cosmological singularities in general relativity \cite{Lifshitz:1963ps, Belinsky:1970ew, Belinsky:1982pk}. They were primarily concerned with the question of the genericity of singularities: are singularities artifacts of our large symmetry assumptions, as one may expect of the singularity in FRW cosmologies? Or would singularities be found in a set of nonvanishing measure in the full space of solutions (if we were only powerful enough to construct them)?

One may ask a similar guiding question in the context of AdS/CFT. Namely, there have been many examples of holographic RG flows in AdS/CFT leading to singular interiors, and the natural question would then be whether this is generic or simply an artifact of sourcing operators that preserve various symmetries, like Poincar\'e symmetry. This is precisely analogous \cite{Shaghoulian:2013qia} to the symmetric FRW cosmologies BKL were worried about. It is this question with which we will be occupied in the remainder of the paper. The singularities we will construct will be timelike instead of the usual spacelike singularities of BKL. Such timelike singularities will allow a cleaner interpretation in terms of boundary renormalization group flow. 

Through several iterations of the work of BKL, which later involved cross-fertilization with the West and the methods of the Penrose/Hawking school, their work came to be known as the BKL conjecture. Roughly stated, the BKL conjecture claims the existence of a generic class of spacelike singularities in general relativity (coupled to matter) which are  approximately ultralocal, oscillatory, and strong. ``Generic" here means the solution is not a set of measure zero in the space of all solutions. ``Ultralocal" means that as one approaches the singularity, in the Hamiltonian formalism, one can replace the partial differential equations governing the initial data with ordinary differential equations. This is possible since the time derivatives dominate over the space derivatives. ``Approximately ultralocal" means this replacement is valid except at isolated points in time which can be modeled as sharp walls. ``Oscillatory" means the solution oscillates, point-by-point, through a family of parameterizable sub-solutions; these ``oscillations" are induced by the sharp walls which break exact ultralocality. ``Strong" means that the metric cannot be analytically continued past the singularity in any continuous way. We will make these notions precise as we go along. Certain types of matter couplings are known to prevent the existence of BKL-type solutions with the above properties, but we will only consider pure gravity. The formalism of section \ref{billiards} makes the inclusion of matter straightforward. 

Generic gravitational phenomena have been re-interpreted in the past few decades in terms of generic field-theory phenomena, primarily through the AdS/CFT duality. The simplest example is a large, finite-temperature black hole in AdS, which corresponds to a field theory placed at finite temperature. Another example is the Hawking-Page transition, which has been connected to a confinement-deconfinement transition in field theory. Yet another example is black holes which are bald at high temperatures yet develop scalar hair at low temperatures, which is related to superconductivity in field theory. The list of such connections is long and some entries are more speculative than others. The generic nature of the BKL singularity suggests that it be included in the pantheon of gravitational phenomena which have been related to field theory phenomena. Of course, the BKL conjecture in all its glory remains unproven, but there exist analytic solutions and numerical simulations which realize various properties of the conjecture.

\subsection{Summary of results}

We will construct a timelike cousin of the BKL singularity and show that it has properties similar to the spacelike version. Specifically, there will be three key results: (1) Such timelike singularities are ultralocal from the viewpoint of a radial Hamiltonian evolution toward the singularity, (2) Chaotic behavior can be shown in $(3+1)$ dimensions, but does not extend simply to higher dimensions, and (3) Imposing causality constraints in the dual field theory  severely limits the types of timelike BKL singularities which can be embedded into AdS/CFT, potentially ruling out the chaotic subclass completely. The first two results are general properties of the singularity independent of AdS/CFT, whereas the third property is inherent to AdS/CFT.   

We then turn to some speculation.  Assuming that such a singularity exists in the interior of an asymptotically AdS solution, we will interpret the oscillatory, chaotic approach to the singularity as the bulk dual of a chaotic renormalization group flow. Since the ultimate infrared is unknown at the classical level and the singularity may be resolved by quantum or stringy effects, the chaotic behavior may only be transient. We will also illustrate that one way to think about the dual field theory is akin to a duality cascade which--unlike the duality cascades considered thus far--has the chaotic bouncing visible at the classical level in the bulk.

{\bf v2}: Since the publication of this paper, it has been brought to our attention that--as concerns Section \ref{oscillatory} about timelike BKL singularities in $3+1$ dimensions--Serge L. Parnovsky beat us to print by a whopping 36 years \cite{PARNOVSKY1980210}. That work overlaps and agrees with our analysis in that section. The author further showed that two other candidate metrics near timelike singularities cannot be upgraded to generic, chaotic solutions (see also \cite{Tomita:1977yg}). In \cite{1985ZhETF..88.1921P, 0264-9381-7-4-008}, the same author considered several less exact solutions and studied the case of null singularities. These papers do not treat the general off-diagonal case, which we are able to do with the formalism of Section \ref{billiards}. While our work was being completed, the author of \cite{Klinger:2015vhx} performed the $(3+1)$-dimensional Iwasawa analysis of Section \ref{billiardsdetail}. That paper attaches physical significance to the symmetry wells in $3+1$ dimensions, which we argue in Section \ref{billiardsdetail} and Appendix \ref{symwells} is incorrect. Our approach resolves the inconsistency of \cite{Klinger:2015vhx} with the results of Parnovsky.\footnote{Thanks to John D. Barrow and Piotr Chrusciel for bringing these papers to our attention.}

\section{BKL summary}\label{recap}
We will briefly review the salient points of the classic analysis of Belinski, Khalatnikov, and Lifshitz (BKL), as represented in the papers \cite{Lifshitz:1963ps, Belinsky:1970ew, Belinsky:1982pk}. A fuller summary is given in appendix \ref{appkl1} - \ref{appmatter}. 

The anisotropic Kasner geometry, a solution of the vacuum Einstein equations, is written as 
\begin{align}
ds^2=-dt^2+t^{2p_1}dx^2+t^{2p_2}dy^2+t^{2p_3}dz^2\,,\label{kasner}\\
 p_1+p_2+p_3=p_1^2+p_2^2+p_3^2=1\,.\label{exps}
\end{align}
BKL's more general ansatz for a singularity at $t=0$ was 
\begin{align}
ds^2=-dt^2+\left(t^{2p_l(\vec{x})} l_i(\vec{x})l_j(\vec{x})+t^{2p_m(\vec{x})} m_i(\vec{x})m_j(\vec{x})+t^{2p_n(\vec{x})}n_i(\vec{x})n_j(\vec{x})\right)dx^idx^j\,,\label{kasnergen}\\
p_l(\vec{x})+p_m(\vec{x})+p_n(\vec{x})=p_l(\vec{x})^2+p_m(\vec{x})^2+p_n(\vec{x})^2=1\,.\label{expsgen}
\end{align}
The full geometry can of course be much messier, but this metric represents the leading terms of the metric in an expansion in $1/t$ near the singularity at $t=0$. For the case where the vectors $l_i$, $m_i$, and $n_i$ are independent of position and point along the axes $x_i$, and the $p_i$ are also independent of position, this reduces to the usual anisotropic Kasner geometry.

In writing down Einstein's equations in an expansion in $1/t$, one finds there are two possibilities. The first possibility is that all spatial gradients are subleading compared to the time derivatives. Thus, the system becomes ``ultralocal." This property is precisely stated in terms of the Hamiltonian formulation of general relativity and is discussed in \cite{Isenberg:1989gq}. In short, ultralocality is the statement that the partial differential equations encoding the evolution of the three-metric and its extrinsic curvature can be replaced with ordinary differential equations asymptotically close to the singularity, due to the subleading nature of the spatial derivatives and spatial curvature terms. There is no such constraint on the Hamiltonian and momentum constraints coming from variations with respect to the lapse and shift, which for ultralocal systems can be partial differential equations. In a generic system it is incorrect to make this approximation, but there are tractable analytic examples where the approximation becomes arbitrarily good as you approach arbitrarily close to a spacelike singularity. One such example is given by the family of polarized Gowdy spacetimes with the spatial topology of a three-torus \cite{Isenberg:1989gq}. Although there is analytic evidence for this portion of the BKL analysis, this situation is not generic, as it is one free function short of a fully generic solution as illustrated in \cite{Lifshitz:1963ps}.

The other possibility, first realized in \cite{Belinsky:1970ew}, is that the approach to the singularity can be oscillatory and approximately ultralocal. In this case, the spatial derivatives \emph{cannot} be neglected, and the evolution of the equations is such that the system can undergo an infinite number of oscillations between different Kasner epochs. This possibility is realized in the Bianchi IX cosmology with non-stiff matter \cite{Andersson:2000cv}. The allowance of oscillatory behavior upgrades the solution to a generic one. Numerical evidence exists for this part of the BKL conjecture as well \cite{thorne}.

 For interesting work embedding the spacelike Kasner singularities \eqref{kasner} - \eqref{exps} in AdS/CFT, see \cite{Engelhardt:2013jda, Engelhardt:2014mea}. For unrelated recent work investigating chaos in black hole backgrounds, see \cite{Shenker:2013pqa, Shenker:2013yza, Roberts:2014isa, Roberts:2014ifa, Shenker:2014cwa, Polchinski:2015cea}. Parameterizing chaos in relativistic systems with Lyapunov exponents  is subtle, due to the sensitivity of this measure to time reparametrizations. For more invariant notions of chaos in e.g. mixmaster universes \cite{Misner:1969hg}, see \cite{Barrow:1981sx, Chernoff:1983zz, Barrow:1984ag, Cornish:1996hx, Cornish:1996yg, Imponente:2001fy}. For work extending the original billiard picture of Misner \cite{Misner:1969ae, Misner} but predating \cite{Damour:2002et}, see \cite{Ivashchuk:1994da, Ivashchuk:1994fg, Ivashchuk:1999rm, Damour:2000hv}. Finally, for a sample of some quantum BKL excursions, see \cite{Ashtekar:2008jb, Kleinschmidt:2009cv, Kleinschmidt:2009hv, Ashtekar:2011ck, Czuchry:2012ad, Czuchry:2014hxa, Mielczarek:2014via, Bergeron:2015lka, Bergeron:2015ppa, Lecian:2013lxa, Ivashchuk:2013jla, Ivashchuk:2013psa}. 


\section{Timelike BKL singularities}
We begin by exhibiting the radial version of the Kasner geometry \eqref{kasner}, which is also a solution of the vacuum Einstein equations in four dimensions:
\begin{align}
ds^2=dr^2&+(-r^{2p_t}dt^2+r^{2p_x}dx^2+r^{2p_y}dy^2)\,,\label{kasnerr}\\
p_t&+p_x+p_y=p_t^2+p_x^2+p_y^2=1\,.
\end{align}
Analogously to \eqref{kasnergen} - \eqref{expsgen}, this radial Kasner geometry can be generalized to
\begin{align}
ds^2=dr^2+&\left(-r^{2p_l(x_\rho)} l_\mu(x_\rho)l_\nu(x_\rho)+r^{2p_m(x_\rho)} m_\mu(x_\rho)m_\nu(x_\rho)+r^{2p_n(x_\rho)}n_\mu(x_\rho)n_\nu(x_\rho)\right)dx^\mu dx^\nu\,,\label{kasnergenr}\\
p_l(x_\rho)&+p_m(x_\rho)+p_n(x_\rho)=p_l(x_\rho)^2+p_m(x_\rho)^2+p_n(x_\rho)^2=1\,.\label{expsr}
\end{align}
Setting the exponents $p_i$ to constants and aligning the coordinate axes with $l_{\mu}$, $m_{\mu}$, and $n_{\mu}$ gives \eqref{kasnerr}. However, as in the spacelike BKL case the role played by this geometry goes beyond just the vacuum Einstein  equations; it will be our ansatz for the leading terms in a $1/r$ expansion of a metric with timelike singularity at $r=0$, coming from a potentially much more general action. As a simple, concrete example of how stress-energy can be neglected near the singularity, let us consider the case of Schwarzschild-AdS, which is asymptotically Kasner even though the full geometry is a solution of Einstein's equations with negative cosmological constant. The metric is given by 
\be
ds_{d+2}^2=-\left(1+\frac{r^2}{\ell^2}-\frac{2M}{r^{d-1}}\right)dt^2+\frac{1}{1+\frac{r^2}{\ell^2}-\frac{2M}{r^{d-1}}}\,dr^2+r^2d\Omega_d^2\,.
\ee
We can take $r\rightarrow 0$ to approach the singularity and get 
\be
ds^2\approx \frac{2M}{r^{d-1}}dt^2-\frac{r^{d-1}}{2M}dr^2+r^2d\Omega_d^2\,.
\ee
Notice that the time coordinate is $r$ if $M>0$ (we are inside a usual black hole horizon), while it is $t$ if $M<0$. We consider $M<0$, zoom in on a flat patch of the sphere, and absorb the factor of $-2M$ into the coordinates to get
\begin{align}
ds^2\approx -\frac{dt^2}{r^{d-1}}+r^{d-1} dr^2+r^2dx_i^2\,.
\end{align}
Defining $r^{d-1} dr^2\equiv d\tilde{r}^2$ gives
\begin{align}
ds^2\approx d\tilde{r}^2-\tilde{r}^{2(1-d)/(d+1)}dt^2+\tilde{r}^{4/(d+1)}dx_i^2\,.
\end{align}
This is a radial Kasner geometry \eqref{kasnerr} with exponents $p_t=(1-d)/(d+1)$, $p_i=2/(d+1)$. We will see that these exponents are ruled out by the causality constraint in section \ref{causality}. Notice that the appearance of the Kasner geometry illustrates that the cosmological constant--interpreted as a matter source--has decoupled near the singularity. 

\subsection{Decoupling near the singularity}
We begin by considering the less generic case of no oscillation, first considered in \cite{Lifshitz:1963ps}. This is the timelike analog of Section \ref{appkl1}. We consider Einstein's equations in vacuum and will assume any matter stress-energy is negligible near the singularity. We work in a synchronous reference system and expand the metric about the singularity as 
\be
ds^2=dr^2+(-a^2l_\mu l_\nu+b^2m_\mu m_\nu+c^2n_\mu n_\nu)dx^\mu dx^\nu=dr^2+h_{\mu\nu}dx^\mu dx^\nu\,,
\ee
where $a$, $b$, $c$, $l_\mu$, $m_\mu$, and $n_\mu$ all depend on $r$ and $x_\mu$. The metric is seen to be Lorentzian by computing the determinant in the basis given by $\{l_\mu,m_\mu,n_\mu\}$, which gives det $g_{\mu\nu}=-(\epsilon^{\mu\nu\rho}l_\mu m_\nu n_\rho abc)^2$. Notice it is in general not possible to align the $x_\mu$ with the directions $l_\mu$, $m_\mu$, and $n_\mu$. 

Defining $\kappa_\mu^\nu=\partial h_\mu^\nu/\partial r$, we can write Einstein's equations in this frame as
\begin{align}
R_r^r&=-\frac{1}{2}\frac{\p \kappa_\mu^\mu}{\p r} -\frac{1}{4}\kappa_\mu^\nu\kappa_\nu^\mu=0 \,,\label{firsteqr}\\
R_\mu^\nu&=-\frac{1}{2\sqrt{|g|}}\frac{\p}{\p r}(\sqrt{|g|}\kappa_\mu^\nu)-^{(3)}R_\mu^\nu\,,\label{secondeqr}\\
R_\mu^r&=\frac{1}{2}\left(\nabla_\nu\kappa_\mu^\nu-\nabla_\mu\kappa_\nu^\nu\right)\,.\label{thirdeqr}
\end{align}
The first equation only has derivatives with respect to $r$. The same goes for the second equation if we can ignore the three-dimensional Ricci tensor built out of $h_{\mu\nu}$.\footnote{The oscillatory behavior, which will be discussed in the next subsection, occurs for precisely the case when you cannot ignore $^{(3)}R_\mu^\nu$ because it has terms that are $\mathcal{O}(r^{-n})$ for $n>2$.} Then these two equations become 
\begin{align}
R_r^r=\frac{a''}{a}+\f{b''}{b}+\f{c''}{c}&=0\,,\\
R_l^l=-\f{(a'bc)'}{abc}=0\,, \qquad R_m^m=-\f{(ab'c)'}{abc}&=0\,, \qquad R_n^n=-\f{(abc')'}{abc}=0\,,
\end{align}
where we have defined projections of the tensor $R_{\mu\nu}$ as e.g. $R_{mn}=R_{\mu\nu}m^\mu n^\nu$. These equations have as solution the anisotropic Kasner geometry \eqref{kasnergenr} - \eqref{expsr}. The final equation \eqref{thirdeqr} can be reduced to three $r$-independent constraints on the exponents and vectors \cite{Lifshitz:1963ps}. As we will see in the next subsection, ignoring $^{(3)}R_\mu^\nu$ requires a single constraint. This additional constraint makes the parameterization a set of measure zero in the full phase space; including it will make the ansatz cover an open set in the space.

The equations \eqref{firsteqr}-\eqref{thirdeqr} can also be recast in the Hamiltonian formalism, where one uses a radial Hamiltonian appropriate for evolution toward the interior. In this framework, the evolution equations are analogous to \eqref{firsteqr} and \eqref{secondeqr}, while the constraint equations are analogous to \eqref{thirdeqr}. Ultralocality is then the statement that the non-radial derivative terms in the \emph{evolution} equations can be ignored asymptotically close to the singularity. This means that the evolution equations break up pointwise, and it allows us to zoom in to different points and track their evolution independently from nearby points. In other words, the set of partial differential equations governing the radial evolution of the three-metric become ordinary differential equations. Even in the case where the term $^{(3)}R_\mu^\nu$ has spatial derivatives which can compete, it turns out that the spatial derivatives act in a rather mild way. Specifically, in a suitable limit (the BKL limit) we can ignore the spatial derivatives except at isolated points in time, which can be treated as sharp transitions. We will refer to this situation as ``approximately ultralocal" and address it in the next subsection.

A potentially puzzling feature of the radial evolution becoming ultralocal is the following: in the usual BKL case of a spacelike singularity, one can appeal to the existence of particle horizons near the singularity as an intuitive explanation for the point-by-point decoupling,\footnote{This was emphasized to us by Gary Horowitz.} whereas there is no such causal picture in the case of radial evolution. Nevertheless, the picture of horizons can be supplanted by the picture, at least in the synchronous gauge within which we work, of the strength of the singularity leading to certain terms in the dynamics dominating over others. This is what happened in the symmetric case above. This picture also works in explaining analogous decoupling in the case of cosmological horizons, in which case there is no singularity but the strength of the expansion replaces the strength of the singularity. Furthermore, the explanation via horizons is not completely satisfactory since the oscillatory scenario we will illustrate in the next section violates the strict decoupling, and this generalized behavior does not have a clear explanation in terms of the behavior of horizons. Interestingly, BKL never referenced the existence of causal horizons to support their decoupling picture, and instead argued directly from the equations of motion, which is what we do as well. We will not spend much time on the inhomogeneous case, but preliminary work suggests that the arguments of \cite{Belinsky:1982pk}, which show that the homogeneous case can capture the inhomogeneous case, can be adapted to our scenario. To be clear, the pointwise split of a generic inhomogeneous situation into homogeneous but anisotropic patches, as seems to happen in the spacelike BKL singularity, is an assumption of our work.

\subsection{Oscillatory behavior}\label{oscillatory}
We can now consider the case where $^{(3)}R_\mu^\nu$ is kept finite. This is the timelike analog of Section \ref{appbkl2}, which we rely on heavily and should be read before reading this section. We write the metric as
\be
ds^2=dr^2+(-a(r)^2l_{\mu}l_{\nu}+b(r)^2m_{\mu}m_{\nu}+
c(r)^2n_{\mu}n_{\nu})dx^{\mu}dx^{\nu}\,,
\ee
where we have singled out $\vec{l}$ as the timelike direction. The corresponding Einstein equations can be obtained by changing $\frac{d}{dt}\rightarrow i\frac{d}{dr} \,,\,a^2\rightarrow -a^2$ in \eqref{eqnsosc}:
\bea\label{eqnsoscr}
-\frac{(a'bc)'}{abc}=\frac{1}{2a^2b^2c^2}[\lambda^2 a^4-(\mu b^2-\nu c^2)^2]\,,\nonumber\\
-\frac{(ab'c)'}{abc}=\frac{1}{2a^2b^2c^2}[\mu^2 b^4-(\lambda a^2+\nu c^2)^2]\,,\nonumber\\
-\frac{(abc')'}{abc}=\frac{1}{2a^2b^2c^2}[\nu^2 c^4-(\lambda a^2+\mu b^2)^2]\,,\nonumber\\
-\frac{a''}{a}-\frac{b''}{b}-\frac{c''}{c}=0\,,
\eea
where the primes now denote radial derivatives. We have defined $\lambda= l^i \epsilon_{ijk}\partial_{x_j}l^k$, $\mu= m^i \epsilon_{ijk}\partial_{x_j}m^k,$ and $\nu= n^i \epsilon_{ijk}\partial_{x_j}n^k$, which can be functions of the non-radial coordinates (we use Latin indices in these expressions to minimize the confusion of using $\mu$, $\nu$, $\lambda$ both as indices and as the parameters just defined). In the previous subsection, we solved \eqref{eqnsoscr} in the limit where $\lambda=\mu=\nu=0$. This is precisely the case where we set $^{(3)}R_\mu^\nu=0$ in \eqref{secondeqr}, as advertised. We will now consider the more general case of $\lambda, \mu,\nu\neq 0$. In this case, a pure Kasner solution of $a(r),b(r),c(r)$ will cease to be valid at some point, since one of the scale functions, say $a(r)$, will be diverging instead of vanishing toward the singularity, due to the fact that one of the kasner exponents $(p_1,p_2,p_3)$ has to be negative. This corresponds to the direction along $l_\mu$ expanding instead of contracting. This will give a big contribution to the right-hand-side of \eqref{eqnsoscr}, and we can capture its effect by keeping the dominant $a^4$ term and tracking how it modifies the Kasner solution. This is elaborated in the spacelike case in Section \ref{appbkl2}. From the discussion there, we know that the oscillatory behavior comes from the distribution of signs of the quartic terms: $\frac{(a'bc)'}{abc}\sim -\lambda^2a^4+\mu^2b^4+\nu^2c^4...$, etc. A sign reverse can potentially change the picture of a particle colliding into an exponential potential wall to a particle sinking into an exponential potential well. However, the double-Wick-rotation only affects the sign of the double-quadratic terms $a^2b^2$ and $a^2c^2$ in the brackets on the right-hand-side of (\ref{eqnsoscr}). As argued in Section \ref{appbkl2}, it is the quartic pieces which are relevant and lead to chaotic behavior; the double-quadratic pieces are subleading and will change the details of the collision, but the asymptotic outcome of the bouncing is unaffected. We therefore conclude that the chaotic behavior is retained. We plot the features of this type of oscillatory behavior in Figure \ref{timelike_curv} in the language of the billiard dynamics of the next section.

\section{Radial billiards}\label{billiards}
A generic formulation of the BKL hypothesis in terms of billiards on hyperbolic space was presented in \cite{Damour:2002et} and reviewed in \cite{Henneaux:2007ej}. The idea is to perform an ADM decomposition and map the gravitational dynamics to the simple problem of a billiard ball bouncing around in a stadium in hyperbolic space. This billiard problem takes place in superspace (the space of metrics), and  the walls of the billiard descend from potentials in the ADM decomposition. The potential walls are generically exponentials, but become infinitely sharp in the BKL limit.  The momentum components of the billiard are related to the Kasner exponents $\left\lbrace p_i\right\rbrace$, so bouncing off of walls transitions you from one Kasner epoch to another by changing the momentum of the billiard. Compact regions in this hyperbolic billiard result in a chaotic billiard.  There are two types of walls in this analysis, known as gravity walls and symmetry walls, which have different origins in the ADM Hamiltonian. In $3+1$ dimensions,  the symmetry walls play a rather benign role and primarily reflect the permutation symmetry between spacelike directions. Beginning with only gravity walls forming a ``big billiard," one can then take a quotient by the symmetry group and obtain a ``small billiard" demarcated by a combination of the symmetry walls and gravity walls (the quotiented and unquotiented dynamics is discussed further in \cite{Damour:2010sz}). This benign role of symmetry walls in $3+1$ dimensions is also clear from the analysis of BKL, in which there are no such walls that appear. It becomes tempting to conclude these are unphysical walls which only arise from the diagonalization required by \cite{Damour:2002et}. However, symmetry walls play a more physical role in higher dimensions. For example, whether or not certain models are chaotic depends on whether or not off-diagonal metric components, and therefore symmetry walls, are included \cite{Demaret:1988sg, deBuyl:2003iob}.

The analysis of \cite{Damour:2002et} included the effects of dilatons and $p$-forms, with a view toward analyzing the BKL conjecture for supergravity theories in general dimensions. Having shown in Section \ref{oscillatory} that pure gravity in four dimensions--with a specific ansatz that restricts off-diagonal components in $\gamma_{\mu\nu}$--has chaotic behavior, one may want to refine the analysis and see how much of the mathematical machinery from \cite{Damour:2002et} carries over to the case of a timelike singularity. The answer is that the machinery can be partially adapted, although in certain cases the formalism breaks down.

We will analyze pure gravity in $d+1$ dimensions. We will find that chaos is preserved for $d=3$ even when we relax the restriction on off-diagonal components, which we had imposed in the previous section. There is only one symmetry wall (as opposed to three symmetry walls in the spacelike BKL case), presumably reflecting the reduced permutation symmetry of the problem (two equivalent spacelike coordinates as opposed to three). For $d>3$, we provide evidence that gravity is not chaotic. For example, in $4+1$ dimensions, we argue that three of the twelve potential gravity walls of the spacelike BKL billiard turn into \emph{wells} due to the timelike nature of one of the coordinates. These wells need to be resolved to know where the dynamics leads, but we will not address this issue here. We will also find a reduced number of symmetry walls, and imposing them is not sufficient to block the gravity wells and restore chaos. We will be sparse with the details of the billiard construction, relying heavily on the beautiful exposition of \cite{Damour:2002et, Henneaux:2007ej}.

\subsection{Billiard in hyperbolic space}\label{billiardsdetail}
Since in our setup one approaches the singularity in a spacelike direction, one needs to perform a radial version of the usual ADM decomposition \cite{Arnowitt:1962hi}. For simplicity we will consider pure gravity. We write a metric of the form 
\be
ds^2=dr^2+\gamma_{\mu\nu} dx^\mu dx^\nu
\ee
for $\gamma_{\mu\nu}$ a $d$-dimensional metric of Lorentzian signature. In terms of the extrinsic curvature $K_{\mu\nu}$ of the metric $\gamma_{\mu\nu}$, we have the radial ``Hamiltonian"
\be
H=\int d^d x \sqrt{\gamma}\left(-(K_{ij} K^{ij} - K^2) - \leftidx{^{(d)}}{R}{}\right)\,
\ee
and Lagrangian
\be
L=\int d^d x \sqrt{\gamma}\left(-(K_{ij} K^{ij} - K^2) + \leftidx{^{(d)}}{R}{}\right)\,.
\ee
Integrations are over all non-radial variables. These expressions differ from the usual ADM construction by an overall sign on the extrinsic curvature pieces. The lower-dimensional Ricci scalar is identified with the potential, while the rest is identified with the kinetics. The relation 
\be
\pi^{\mu\nu}\equiv \frac{\delta L}{\delta \dot{\gamma}_{\mu\nu}}=-\sqrt{\gamma}\,(K^{\mu\nu}-\gamma^{\mu\nu}K)
\ee
for the momentum conjugate to $\gamma_{\mu\nu}$ has been used to Legendre transform between the Lagrangian and the Hamiltonian. $\dot{\gamma}_{\mu\nu}$ refers to differentiation with respect to the radial coordinate.

This overall sign flip on the extrinsic curvature pieces is the only difference from the case in \cite{Damour:2002et}. It leads to a (truncated) $d$-dimensional Dewitt supermetric \cite{DeWitt:1967yk} of signature $(+,-,-,\dots,-)$. The billiard problem is defined as the dynamics of a particle in this auxiliary superspace, whose metric can be written as 
\be
d\sigma^2=-\left(\sum_{i=1}^d (d\beta_i)^2+\left(\sum_{i=1}^d d\beta_i\right)^2\right)=d\rho^2-\rho^2d\mathcal{H}_{d-1}^2
\ee
for the unit hyperboloid $\mathcal{H}_{d-1}$. The transformation between $\beta$ variables and the hyperbolic variables is given by $\beta^\mu = \rho \gamma^\mu$, with $\gamma^\mu \gamma_\mu = 1$. The coordinates $\gamma^\mu$ parameterize the sheet of the hyperboloid with $\beta_\mu \beta^\mu > 0$. 

In the case of a spacelike singularity, one is able to map the dynamics to a free particle on a unit hyperboloid with metric $d\mathcal{H}_{d-1}$. In our case, the Hamiltonian flips sign, so the free particle has a wrong-sign kinetic term, but the metric signature also flips sign. The kinematics is thus invariant under these replacements, and we again recover a hyperbolic billiard.

To analyze the Kasner epoch transitions, which in the reformulated hyperbolic billiard problem correspond to the particle bouncing off of walls in hyperbolic space, we need to analyze the generic case of a metric with off-diagonal components. This leads to a ``centrifugal" or ``symmetry" wall, which is the contribution of off-diagonal pieces to the free-particle kinetic piece $\left(-(K_{ij} K^{ij} - K^2)\right)$, and a ``curvature" wall, which is the contribution of the three-dimensional Ricci scalar. In other words, one splits the Hamiltonian as
\be\label{eq:walls}
H=H_0+V_S+V_G; \qquad V_G\equiv -\sqrt{\gamma}\; \leftidx{^{(d)}}{R}{}, \quad V_S=-\frac{1}{2}\sum_{a<b} \f{A_b^2}{A_a^2}(P^j_b\mathcal{N}^a_j)^2\,.
\ee
The piece $V_S$ is written in terms of an Iwasawa decomposition of the metric $\gamma=\mathcal{N}^TA^2\mathcal{N}$ with $A_b=e^{-\beta^b}$ for $b$ a spatial index and $A_t=ie^{-\beta^t}$.\footnote{For general Lorentzian metrics an Iwasawa decomposition is not guaranteed to exist, so we are considering only the sub-class where they do exist. The breakdown of this formalism which we are about to encounter may be related to this assumption.} This is a diagonalization into a  Kasner-type metric $A^2_{ab}=e^{-2\beta^b}\eta_{ab}$ (no sum intended). $P^j_b$ is the momentum conjugate to $\mathcal{N}^b_j$. Details for the spacelike case can be found in \cite{Damour:2002et}. In the BKL limit the pieces $V_S$ and $V_G$ become a sum of theta-function walls of the scale variable $\beta_i$'s, which provide an infinite potential that bounds the kinematics to remain in one side of the walls. 

Let us begin by analyzing the symmetry walls, which come from the extrinsic curvature pieces. But first let us set up some expectations. In $3+1$ dimensions, insofar as symmetry walls tend to reflect the permutation symmetry of the coordinates in the three-metric, we should expect to find a reduced number of symmetry walls compared to the spacelike BKL case. This is because the symmetric permutation group reduces from $S_3$ to $S_2$ due to the timelike coordinate. If it exactly mirrored the symmetries, we would expect to go from three walls to one wall, which is precisely what we find. In higher dimensions, where the symmetry walls play a more physical role, there will not be such an obvious counting, as already shown in \cite{Demaret:1988sg, deBuyl:2003iob}. Nevertheless, we expect some amount of reduction in the number of walls. 

Now to the analysis. Performing a naive analytic continuation of the symmetry walls in the spacelike BKL singularity (which we will soon find is incorrect), we find
\be
V_S=-\rho^2\sum_A c_A e^{-\rho w_A(\beta)}\underbrace{\longrightarrow}_{\rho\rightarrow \infty} -\sum_{a<b}(-1)^n\Theta(-w_{ab}(\beta))\;;\qquad w_{ab}(\beta)\equiv \beta^b-\beta^a
\ee
where $n=1$ if $a$ or $b$ is the timelike coordinate $t$ and $n=0$ otherwise. The $c_A$ coefficients are independent of the phase space variables $\beta$ and $\pi_\beta$. The step function is defined as 
\begin{align}
\Theta(x)= \begin{cases}
               0, \qquad &\textrm{if }x<0\\
               \infty, \qquad &\textrm{if }x>0
            \end{cases}
\end{align}
Let us now consider the gravity walls. For an Iwasawa basis $\gamma= \Sigma\; A_i^2\;\theta_i^2$,
\be
d\theta^a = -\frac{1}{2}\, C^a_{bc} \, \theta^b \wedge \theta^c
\ee
we get a Ricci scalar given by 
\be\label{gravitywall}
R=-\frac{1}{4} \sum_{a\neq b\neq c\neq a} \frac{A_a^2}{A_b^2 A_c^2} (C^a_{bc})^2+\textrm{subleading}\,,
\ee
where ``subleading" means pieces that become irrelevant walls in the BKL limit.\footnote{Irrelevant walls refers to potential walls which are completely hidden behind other walls; it is clear such objects do not affect the dynamics in the BKL limit.} Notice that for $D=4$ every term in the sum always includes $A_t=ie^{-\beta^t}$, which squares to give a minus sign. This does not happen in the spacelike BKL case. The gravity walls become, in an appropriate gauge, 
\be
V_G\underbrace{\longrightarrow}_{\rho\rightarrow \infty} -\sum_{a\neq b\neq c\neq a}\Theta(-\alpha_{abc}(\beta))\;; \qquad \alpha_{abc}(\beta)\equiv2\beta^a+\sum_{e\neq a,b,c}\beta^e\,.
\ee

For pure gravity in four dimensions, we see that $H=H_0+V_S+V_G$ flips the overall signs in the $H_0$ and $V_G$ pieces (compared to the spacelike singularity case), whereas $V_S$ is a sum of three terms, only one of which flips signs. To get the correct kinematics, we fix the wrong sign kinetic term $H_0$ by negating again $V_S$ and $V_G$. Thus, instead of three gravity walls and three symmetry walls in the hyperbolic billiard (which is what happens in the case of a spacelike singularity), it seems that there are three gravity walls, one symmetry wall, and two symmetry wells. However, this analysis is incorrect, as can be guessed by the fact that symmetry walls morphing into wells is in tension with their benign origin in $3+1$ dimensions as implementers of a symmetry group. 

The resolution to this conundrum is as follows. Technically, the Iwasawa coordinate transformation of \cite{Damour:2002et} is becoming singular, as can be seen by working in a different coordinate frame. This is what is signaled by an infinite symmetry ``well." We will re-solve these equations in a different (nonsingular) coordinate frame and see what becomes of the infinite symmetry wells in a representative example (Bianchi IX) in appendix \ref{symwells}. We will find that they become \emph{finite} symmetry wells, and the billiard ball continues its free flight after passing through.

\begin{figure}
\center{\includegraphics[scale=.33]{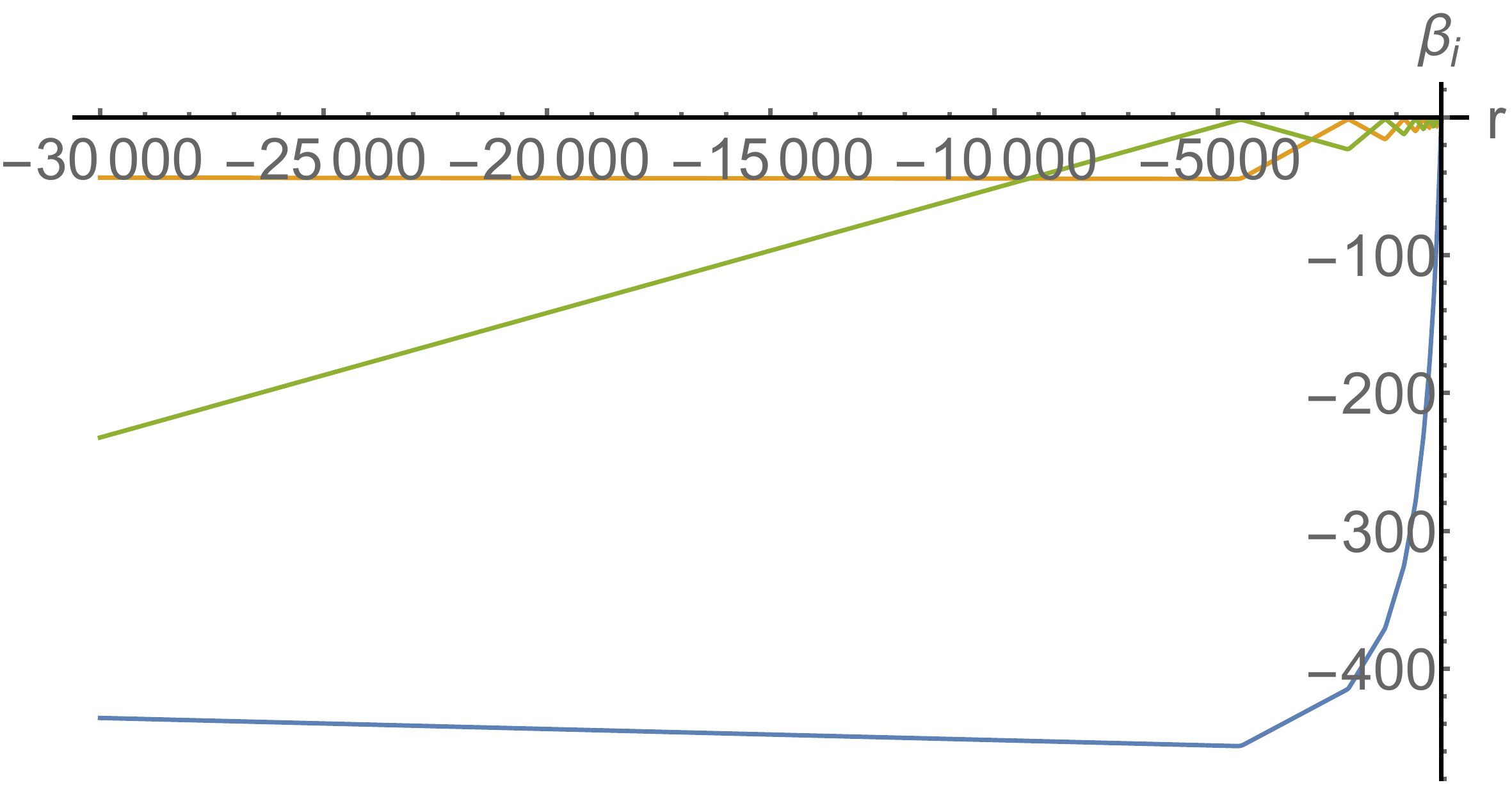}\qquad \includegraphics[scale=.325]{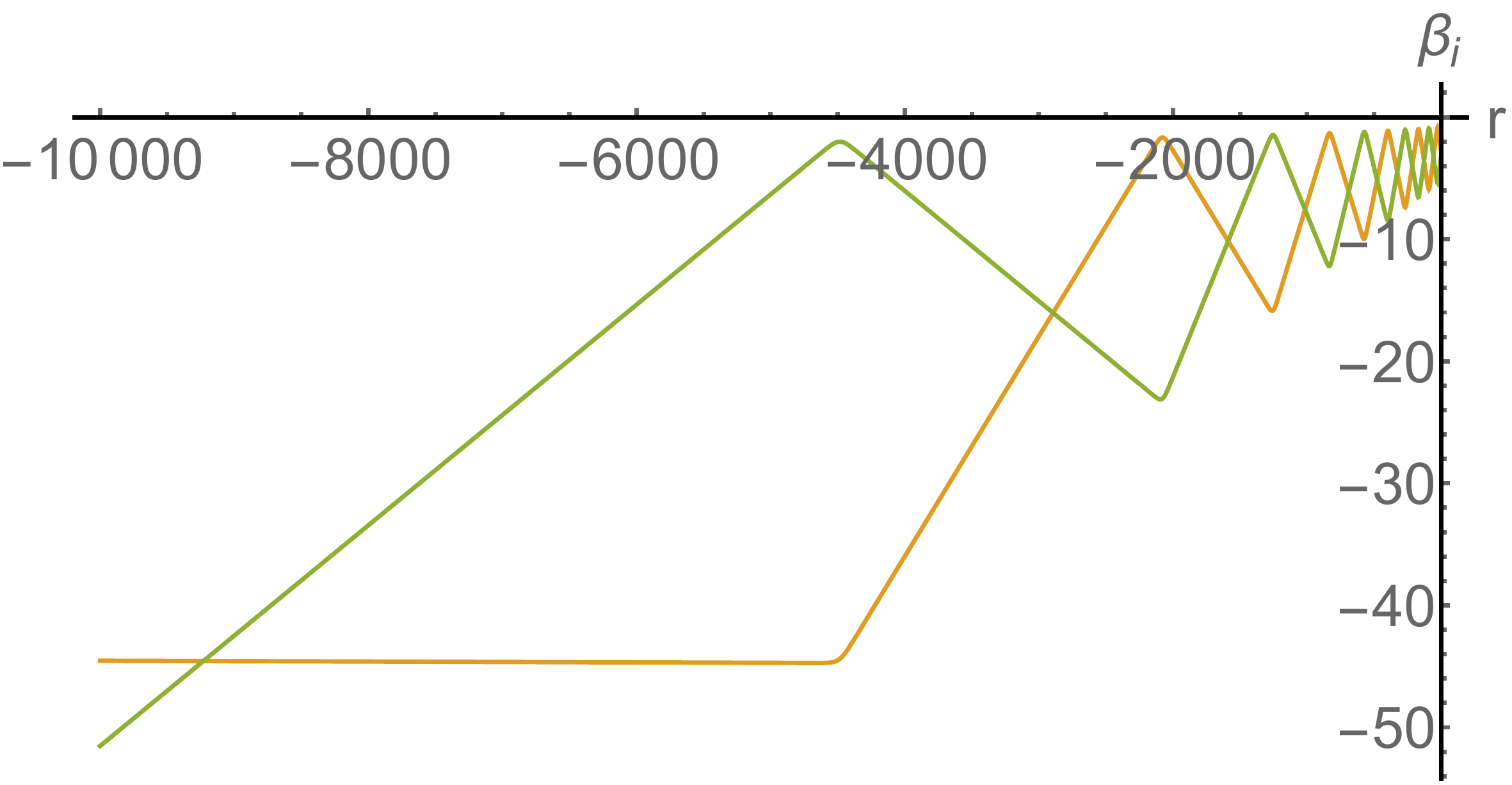}}
\caption{We have plotted the three Iwasawa exponents $\beta_i$ in the absence of symmetry walls, and the evolution toward the singularity is in the negative $r$ direction. The plot on the right zooms closer to the origin and ignores the blue curve for illustrative purposes. This shows cleanly the oscillatory dynamics in this regime. Notice also that the ordering of the Iwasawa exponents are not preserved in the absence of symmetry walls. }\label{timelike_curv}
\end{figure}

If we restrict to only gravity walls (the unquotiented dynamics), we have a simple form of chaotic behavior illustrated in Figure \ref{timelike_curv}. This case is equivalent to the case of a spacelike singularity, and in the language of \cite{Damour:2010sz} it corresponds to the ``big billiard" considered by BKL. 
In four dimensions, the dynamics can be summarized as follows: $(\beta_1,\beta_2,\beta_3)$ will evolve freely along the radial direction with ``momenta" equal to the Kasner exponents $p_i$:
\bea
p_i&\in & \left\lbrace \frac{-u}{u^2+u+1}, \frac{u(u+1)}{u^2+u+1}, \frac{u+1}{u^2+u+1}\right\rbrace,\;u>0\,, \nonumber\\
\beta_i(r)&=&-p_i \log(r)\,.
\eea
Notice that both $u$ and $1/u$ correspond to the same Kasner exponents. The free motion is confined by the gravity walls to be within the region $\beta_i\geq 0$. One of the momenta $p_i$ is negative, which will drive $\beta_i$ towards the boundary of the region. Upon colliding into the boundary, the dynamics will change the trajectory by changing the momenta, and hence the Kasner exponents, in the following way:
\be
p_i\rightarrow -\frac{p_i}{1+2p_i}\,,\qquad p_{j\neq i}\rightarrow \frac{p_j+2p_i}{1+2p_i}
\ee
In terms of the parameter $u$ that characterizes the Kasner exponents, the bouncing dynamics correspond to following rule: if $u>1$, then the bounce sends $u$ to $u-1$; if $0<u<1$, then the bounce sends $u$ to $1/u-1$. From this we see that for any irrational value of $u$, the bouncing dynamics will continue indefinitely. Rational values of $u$ which appear to terminate after a finite number of bounces are finely tuned and appear only as an artifact of the simplifying assumptions we have made.


In the case where we can have off-diagonal pieces in the metric, the analysis in appendix \ref{symwells} shows that we obtain one symmetry wall, corresponding to the single pair of spacelike exponents. Adding in this symmetry wall corresponds to a quotient of the big billiard which gives what we will call the ``small timelike billiard." This billiard is three times the size of the small spacelike billiard of \cite{Damour:2002et} and is discussed in section \ref{coxeter}. 
 

In $d+1$ dimensions it is less clear what is to become of the symmetry walls which  naively continue into wells. Let us for the moment assume it mimics the case of $3+1$ dimensions, meaning they become finite wells. Altogether, we will have $d(d-1)/2-(d-1)$ symmetry walls (including irrelevant ones), $d-1$ finite symmetry wells, and $d(d-1)(d-2)/2$ gravity terms, of which $(d-1)(d-2)(d-3)/2$ are wells and the rest are walls. For example, in $4+1$ dimensions one will have nine gravity walls and three gravity wells. Imposing all of them would give a closed simplex with twelve faces, although three of the faces are wells that allow the billiard ball to escape. Quotienting by the permutation symmetry group of the non-radial spatial coordinates would reduce you to a simplex with five faces: two gravity walls, one gravity well, and two symmetry walls. The fact that the gravity wells did not disappear under the quotient is a consistency check, as the small billiard should encode the physical features of the large billiard. (Notice that imposing symmetry walls is not equivalent to taking the quotient by the permutation symmetry group, as for $d>3$ these are not in one-to-one correspondence. As shown in \cite{Demaret:1988sg, deBuyl:2003iob}, imposing symmetry walls can close an open billiard and restore chaos.) Just as we passed to the orthogonal decomposition in appendix \ref{symwells} to deduce what is to become of the infinite symmetry wells in the Iwasawa decomposition, we again need to do such an analysis for the infinite gravity wells. However, these infinite gravity wells remain in the orthogonal decomposition. Although it may be possible to find a frame in which the gravity wells can be handled, it seems much more likely that the gravity wells are signaling a breakdown of a BKL limit and (modulo our assumptions) an absence of generic chaos in higher dimensions in this particular paradigm.

Although chaos does not seem to be generic in higher dimensions, we can still imagine engineering chaotic solutions. This would correspond to fine-tuning the initial conditions or some other parameters of the problem. For example, we can fine-tune the off-diagonal components of the metric $\gamma$ to take on special values. These special values should be such that the structure constants $C^a_{bc}$ in the Iwasawa frame, which enter into the coefficients of the gravity walls/wells as in equation \eqref{gravitywall}, vanish anytime $a$, $b$, and $c$ are all spacelike coordinates. The indices $a$, $b$, and $c$ being spacelike implies that $C^a_{bc}$ is the coefficient of a gravity well instead of a gravity wall, and the vanishing of all these $C^a_{bc}$ would enforce the vanishing of the gravity wells. In $4+1$ dimensions, for example, the rest of the walls are sufficient in forming a closed billiard. 

\subsection{Timelike vacuum billiard and arithmetic chaos}\label{coxeter}
\begin{figure}

\begin{minipage}[h]{0.5\linewidth}\hspace{1.2cm}
\includegraphics[scale=.22]{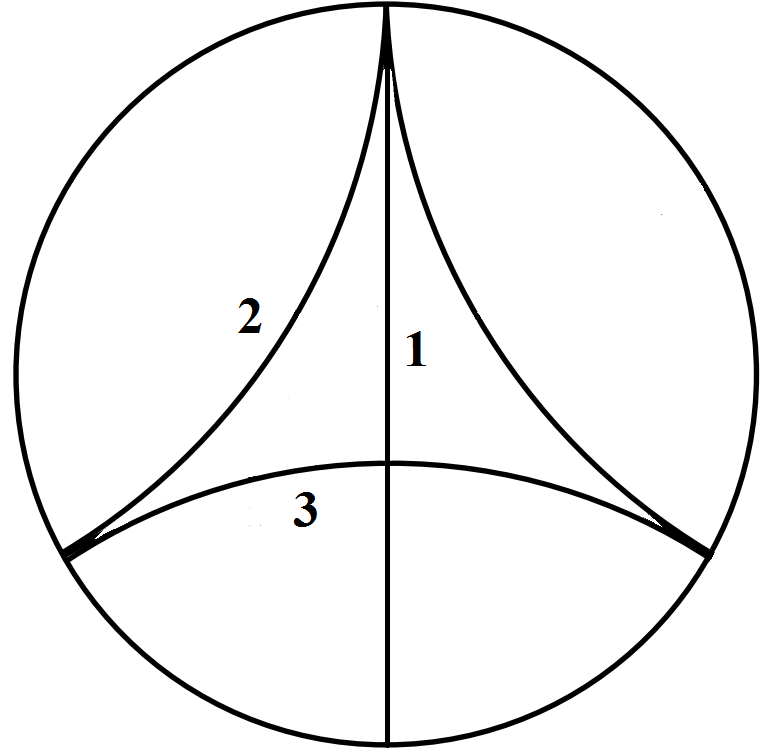}
\label{fig:figure1}
\end{minipage}
\hspace{0.5cm}
\begin{minipage}[b]{0.4\linewidth}
\centering
\begin{tikzpicture}\vspace{-3cm}
 \tikzset{LabelStyle/.style =   {                                text           = black}}

\tikzstyle{line} = [draw=red, -latex']
  \node[fill=red!20,draw,circle] (n1) at (0,0) {1};
  \node[fill=red!20,draw,circle]  (n2) at (2,0)  {2};
  \node[fill=red!20,draw,circle]  (n3) at (4,0)  {3};

 \draw[-, >=latex, color=red!80,double=red] (n1) to node[LabelStyle,above=.2cm]{{\large $\infty$}} (n2) ;
 \draw[-, >=latex, color=red!80,double=red](n2) to node[LabelStyle,above=.2cm]{{\large$\infty$}} (n3);
\end{tikzpicture}
\label{fig:figure2}
\end{minipage}
\caption{In the case of a timelike singularity, we have three gravity walls and one symmetry wall, making up the small timelike billiard on the left-hand figure. This should be compared with Figure \ref{domains}. The Coxeter diagram representing the symmetry of reflections in this billiard is given on the right. The numbered nodes correspond to the reflection walls in the left-hand billiard.}\label{timelike_curv_and_sym}
\end{figure}
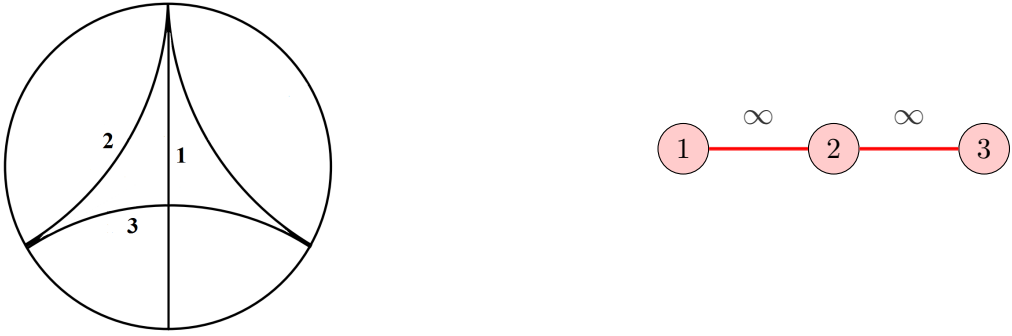

As illustrated in the case of Bianchi IX in appendix \ref{symwells}, the symmetry walls which come from a diagonalization of the metric are harmless. 
For the case of vacuum gravity in $3+1$ dimensions, imposing our single symmetry wall gives the billiard in Figure  \ref{timelike_curv_and_sym}. This billiard is three times the size of the small spacelike billiard of \cite{Damour:2002et}. The angles made up by the intersections of the relevant walls are $0$, $0$, and $\pi/2$.  The reflections of the billiard ball from the walls of this small timelike billiard can be represented by a rank-$3$ Coxeter group with generators $s_1$, $s_2$, and $s_3$ which square to the identity and satisfy the relations $(s_1 s_2)^\infty=1$, $(s_2 s_3)^\infty=1$, and $(s_1s_3)^2=1$. The power $\infty$ is a convention in Coxeter theory which corresponds to the case where there is no power of the product of generators which equals the identity. This leads to the Coxeter diagram in Figure \ref{timelike_curv_and_sym}.

Recall that the spacelike vacuum billiard corresponds to PSL$(2,\mathbb{Z})$ extended by a reflection generator, which altogether gives PGL$(2,\mathbb{Z})$. In our case we begin with the infinite-order Hecke group $H_\infty$, which is the same as the modular group except $T$ acts as a translation by two units instead of one unit. This is then extended by a reflection generator. More precisely, we begin with $S(z)=-1/z$ and $T(z)=z+2$ acting on a complex coordinate $z$ on the upper-half plane. We define the generators of the extended Hecke group as $s_1(z)=-\bar{z}$, $s_2(z)=2-\bar{z}$, and $s_3(z)=1/\bar{z}$. 

We can say more about the Coxeter group corresponding to our billiard. It is hyperbolic and arithmetic. Indeed, we will soon see that hyperbolicity is a necessary requirement for chaos. For a Coxeter diagram that contains no closed cycles, the corresponding group is crystallographic if the link weights are drawn from the set $\{3,4,6,\infty\}$ \cite{coxeter}. This proves that our Coxeter group is crystallographic. Furthermore, crystallographic Coxeter groups correspond to Weyl groups of Kac-Moody algebras, although the Kac-Moody algebra is not uniquely identified. Our small timelike billiard region is known as the ``fundamental Weyl chamber" and the billiard bounces are known as ``Weyl reflections" in this context. Hyperbolicity of the crystallographic Coxeter group implies that it is the Weyl group of some hyperbolic Kac-Moody algebra \cite{Henneaux:2007ej}. In fact, it is the same rank-3 Kac-Moody algebra discussed in a different context in \cite{deBuyl:2003iob}. It has Cartan matrix 
\be
\left( \begin{array}{ccc}
2 & 0 &-2 \\
0 &2 & -2 \\
-2 & -2 & 2 \end{array} \right),
\ee
and is a subalgebra of $AE_3=A_1^{\wedge\wedge}$, which represents the small spacelike billiard.

\section{Timelike BKL singularities in AdS/CFT}

In this section we consider a few aspects of the embedding of these timelike BKL singularities within AdS/CFT. We are unable to conclusively rule out or rule in the existence of such singularities, although we will derive some general constraints. According to \cite{Gubser:2000nd}, a sufficient condition for the acceptability of timelike or null singularities is that the field theory dual can be put at finite temperature, i.e. there exists a near-extremal generalization of the bulk geometry in the form of a black hole which masks the singularity. This provides an IR cutoff which can be taken away in a $T\rightarrow 0$ limit. In such a context, our results would apply to the $T\rightarrow 0$ geometry which maintains a naked singularity. Even for the radial Kasner geometries which we begin with, we have not succeeded in constructing black brane geometries which cloak the singularity. A more useful criterion in our case is that of \cite{Kleban:2001nh}, which demands that two points that are spacelike separated along the boundary metric are also spacelike separated through the bulk. More generally, there should not be any time advance of geodesics passing through the bulk, since this would imply acausal behavior of the boundary field theory.\footnote{Using this causality criterion to evaluate the acceptability of singularities in AdS/CFT was first suggested in \cite{Bak:2004yf}.} The usual theorem covering this situation in AdS/CFT, due to Gao and Wald \cite{Gao:2000ga}, does not apply to our geometries due to the naked singularity. Nevertheless, the criterion of \cite{Kleban:2001nh} will prove useful in ruling out the class of Kasner geometries \eqref{kasnerr} with negative timelike exponent, which includes the case of negative-mass Schwarzschild. Nakedly singular geometries are often dismissed on the basis of cosmic censorship, but this is not very compelling because it is not clear how this conjecture should arise directly from the dynamical equations. (See e.g. \cite{Hubeny:2004cn, Hertog:2004gz, Alcubierre:2004sp} for some work on the cosmic censorship conjecture in AdS spacetimes.)

Proceeding with the assumption that such singularities can exist,  we will then discuss the geometrization of RG flow and show that extremal surfaces do not encounter any barriers that prevent one from probing the BKL region in the deep IR.  We will also briefly address higher curvature corrections in the context of AdS/CFT.

The rest of this section is rather speculative. We will adopt the point of view that these singularities indeed represent IR phases of dual field theories (although not the ultimate IR, for which resolution of the singularity would be needed) and discuss implications. This flight of fancy will culminate in Section \ref{cascades}, where we connect our timelike BKL billiard analysis to the chaotic duality cascades discussed in \cite{Franco:2004jz, Fiol:2002ah}.

Arguments like those of \cite{Luty:2012ww} ruling out infrared chaos are not in tension with ours, because (a) they are for Lorentz-invariant flows, while our flows break Lorentz invariance, and (b) we do not have any comments about the ultimate infrared, which would require singularity resolution.

\subsection{Causality constraint}\label{causality}
The authors of \cite{Kleban:2001nh} proposed a causality constraint for the bulk spacetime to admit a holographic dual. The idea is that information should not travel more quickly through the bulk than through the boundary. This is formalized and proven by the Gao-Wald theorem \cite{Gao:2000ga}, but the presence of the singularity violates the assumptions of the proof and requires an explicit check to be performed.

For simplicity let us focus on the propagation of null quanta. We pick two spatial points on the boundary separated by $\Delta x$ and choose a null path $P$ in the boundary connecting the two points. For any null path $Q$ in the bulk also connecting the two points, we need to satisfy the constraint
\be\label{eq:causality}
\int_Q dt \ge \int_P dt\,.
\ee 
In other words, we need to ensure that the final spacetime point of path $P$ causally precedes the final spacetime point of the path $Q$ as long as they begin their journeys at the same initial spacetime point. Due to coordinate invariance of causal relations, we will simplify the analysis by using the coordinate time $t$ to determine the causal ordering.

Now let us take a particular Kasner epoch:
\be
ds^2 = -r^{2p_t} dt^2 + dr^2 + r^{2p_i} dx^2_i\,.
\ee  
We will work in the framework of local holography and study the causality constraint from the point of view of an effective field theory living at the slice $r=r_*$ of the particular Kasner epoch. We will comment on the validity of such an approach below.

We argue that the causality constraint is satisfied as long as $p_t > p_i$ for all $i$. Consider an arbitrary null curve:
\bea
-r^{2p_t} dt^2 + dr^2 + r^{2p_i} dx^2_i=0\implies r^{-2p_t}(dr/dt)^2+r^{2(p_i-p_t)}(dx^i/dt)^2 =1 \,.
\eea
This implies
\be
v_i^{bulk}(r)\equiv \frac{dx^i}{dt}=r^{p_t-p_i}\left(1-r^{-2p_t}v_r^2\right)^{1/2}\le r^{p_t-p_i}\,.
\ee
Since the boundary path is nailed down at $r=r_*$, we have
\be
v_i^{bdry}(r)=r_*^{p_t-p_i}\,.
\ee
The bulk path satisfies $r(t)\leq r_*$ along the path, so we have $v_i^{bulk}(r) \leq v_i^{bdry}(r)$ as long as $p_t>p_i$. Integrating this relation along the paths shows that the bulk path will always lag behind the boundary path.


A proper analysis of causality would require a global solution that connects the Kasner region to the boundary $AdS_4$. Lacking such solutions to analyze in general, we consider a thin-wall model. Treating the location of the thin wall as a free parameter to try and get the strongest constraint possible, one can see that radial null curves minimize the boundary time $t$ by staying solely in the Kasner region. This means that placing the wall near the AdS boundary will lead to the most time advance.\footnote{ Regardless of where we put the brane, the part of the null curve in the AdS region will not lead to time advance, so it is sufficient to consider the region in the Kasner geometry.} Then we only need to track the constants which arise from matching metrics across the brane. These constants do not affect our results. An explicit example which corroborates this argument is that of AdS-Schwarzschild spacetimes, which satisfy the causality constraint for positive mass yet violate it for negative mass. One can check that working in the full geometry will reproduce the results above, which are interpreted as coming from zooming into the Kasner region.

It is natural to guess that in a generic oscillatory scenario, most of the volume will be contained in regions where $p_t \ngtr p_i$ for some $i$, which would lead to causality violations for appropriately chosen bulk null curves. At the level of our analysis, however, it is not possible to draw any firm conclusions about the causality constraints on oscillatory timelike singularities. 

\subsection{Probing the IR}
Another natural question to consider is whether the boundary field theory has access to the physics of the deep IR. It was shown in \cite{Engelhardt:2013tra} that extremal surfaces anchored at the AdS boundary cannot penetrate hypersurfaces with nonpositive extrinsic curvature (under an additional mild assumption). For the metric ansatz of the form
\be
ds^2=-a(r)^2 dt^2 + dr^2 +b(r)^2 dx^2 + c(r)^2 dy^2\,,
\ee 
one can compute the extrinsic curvature $K=K^\mu_\mu$ for a constant $r$ slice to be $K=\frac{1}{2}\left(\frac{a'(r)}{a(r)}+\frac{b'(r)}{b(r)}+\frac{c'(r)}{c(r)}\right)$.  For a Kasner geometry, one has $K=\frac{p_1+p_2+p_3}{2r}=\frac{1}{2r}>0$. Thus, an oscillatory BKL geometry, with different Kasner patches at different points in spacetime, also has positive extrinsic curvature radial slices. This is true as long as we pick the radial slices away from the transition regions due to gravity wall (for the moment we only the diagonal billiards), which we previously modeled as infinitely sharp walls.  To resolve the details of the wall and check the extrinsic curvature for radial slices within such a wall, we will focus on the homogeneous Bianchi VII geometry, which can drive oscillatory dynamics:
\be
ds^2=dr^2+(-a(r)^2l_{\alpha}l_{\beta}+b(r)^2m_{\alpha}m_{\beta}+
c(r)^2n_{\alpha}n_{\beta})dx^{\alpha}dx^{\beta}\,.
\ee
The vectors $(\vec{l},\vec{m},\vec{n})$ can be identified with the Killing vectors generating the local isometries. The extrinsic curvature of a constant-$r$ slice in such a geometry also takes the form
\bea
K &=& \frac{1}{2}\left(\frac{a'(r)}{a(r)}+\frac{b'(r)}{b(r)}+\frac{c'(r)}{c(r)}\right)=\frac{1}{2}\frac{(abc)'}{abc} \nonumber\\
&=& \frac{1}{2}\partial_\tau (\alpha +\beta +\gamma) e^{\alpha+\beta+\gamma}\,.
\eea
Plugging in the analytic bouncing solutions (\ref{eq:bouncing-solution}), we find that 
\be
K = \left(\frac{e^{2(1+p_1)\tau}\cosh (2p_1(\tau-\tau_0))}{-8p_1}\right)^{1/2}(1+p_1+p_1\tanh(2p_1(\tau-\tau_0)))\,.
\ee
We see that $K$ remains non-negative for $p_1>-\frac{1}{2}$, which is always the case for BKL exponents. 



\subsection{Radial evolution as RG flow}
We argued that the metric \eqref{kasnergenr} represents a generic fate of evolving Einstein's equations radially inward from the asymptotic AdS boundary. However, one would like to extract some statement about the behavior of the dual field theory. The basic assumption which will allow us to do this is that boundary RG flow is geometrized as evolution in some direction toward \emph{or} away from the singularity. We shall assume that it is the usual radial evolution toward the singularity which represents a field theory RG flow to the IR, but we emphasize that this level of specificity is not necessary for our conclusions. The connection between bulk radial evolution and boundary RG flow (including its shortcomings) is well-understood for the case of AdS-like geometries. The most basic intuition for this is to analyze energies as measured by an asymptotic observer. In the AdS case, with metric 
\be
ds^2=\f{-dt^2+dz^2+dx^2+dy^2}{z^2}\,,
\ee
one argues that the warp factor leads to energies measured in the field theory as $E=E_{proper}/z$, which diverges toward the boundary $z\rightarrow 0$ and vanishes toward the horizon $z\rightarrow \infty$. This identification is meaningful primarily in a coarse-grained sense, where the radial evolution is considered on super-AdS scales. 

In our case, the geometry is anisotropic with different warp factors in every direction. Nevertheless, the property $\sqrt{|g|}\sim r$ is satisfied by the Kasner geometries, i.e. the volume decreases linearly with $r$ as you approach the singularity at $r=0$. For the generalized Kasner geometries this holds pointwise and is maintained through the bounces, which simply correspond to swapping exponents. Thus, our interpretation is even applicable in the chaotic case, where the expanding and contracting directions keep swapping. On average, assuming the bounces are uniformly distributed along the various directions, any $g_{\mu\mu}$ component (no sum intended) will be decreasing on average. For the non-chaotic case our interpretation breaks down in the case where $g_{tt}$ is always increasing toward the singularity, which can be arranged as long as $g_{xx}$ and $g_{yy}$ are decreasing. This is precisely the case we have ruled out in Section \ref{causality} by causality constraints, so we need not worry about it. Of course, on the small scales on which we expect our chaos to take place, it is less clear what the dual field theory picture should be, and it is expected that the matrix degrees of freedom are crucially involved.  

Many of the RG flows considered in AdS$_{d+2}$/CFT$_{d+1}$ maintain $(d+1)$-dimensional Poincar\'e invariance \cite{Freedman:1999gk, Gubser:2000nd, Brandhuber:1999hb, Girardello:1999hj}. This is analogous to $(d+2)$-dimensional cosmologies which maintain $(d+1)$-dimensional rotational and translational invariance. The latter are of FRW type whereas the former are of the general scale-covariant type (and become the ``hyperscaling-violating" type when the conformal factor is pure power law). These geometries are too symmetric to fall into the BKL class; indeed, the FRW-type singularity is precisely what motivated Belinski, Khalatnikov, and Lifshitz to search for more generically parameterized curvature singularities.

\subsection{Holographic implication of bulk ultralocal behavior}
The case of strict ultralocal behavior in the bulk, without oscillatory bounces, seems to indicate a dual renormalization group flow which breaks up pointwise. In other words, the IR phase varies point by point, governed by the various Kasner phases in the bulk. Without a microscopic bulk-boundary map we cannot say more about the specific nature of the ultralocal RG flow of the boundary theory.

\subsection{Discrete scale covariance}
Notice that there is a finely-tuned limit of the billiard problem which gives non-chaotic trajectories. This is just the situation where one aims the initial angle of the ball such that it traces out closed trajectories; i.e. after a finite number of bounces the Kasner exponents return to themselves. For example, a set of Kasner exponents with $u=\frac{1}{2}(3+\sqrt{3})$ will return to itself after every 3 bounces against gravity walls. Although the Kasner exponents have returned to themselves, one has moved closer to the singularity after these bounces. This precisely realizes a \emph{discrete scale covariance}. Discrete scale \emph{invariance} in the bulk is what would have corresponded to limit cycles of the renormalization group flow in the boundary theory.\footnote{One may be able to engineer such a situation by considering the scale-invariant version of Kasner geometries, considered by \cite{Nakayama:2009gi}, and trying to obtain analogous bouncing regimes; for other work attempting to realize limit cycles holographically, see \cite{Nakayama:2011zw, Balasubramanian:2013ux}.} We have instead found what is analogous to the hyperscaling-violating cousins of these limit cycles.

\subsection{Connection to duality cascades}\label{cascades}
We now turn to some interesting work in \cite{Franco:2004jz} that constructs chaotic duality cascades in field theory and interprets them as chaotic renormalization group flows. The point of this section is merely to illustrate a field theory example which has some of the features of our bulk solutions. Connecting the two pictures is, at present, an illustrative free association.

 Let us begin with a brief review of the Klebanov-Strassler flow \cite{Klebanov:2000hb} and its interpretation as a ``duality cascade." One begins with $N$ D3-branes and $M$ fractional D3-branes probing a conifold singularity. The dual gauge theory is described by an $SU(N)\times SU(N+M)$ gauge group, with four chiral bi-fundamentals:  $A_1, A_2$ in the $(N+M, \bar{N})$ representation, and $B_1,B_2$ in $(\overline{N+M}, N)$ representation, coupled via a superpotential $W=\lambda tr(A_iB_jA_kB_l)\epsilon^{ik}\epsilon^{jl}$. For $M>0$, the theory is not superconformal, the gauge couplings $(g_1,g_2)$ of the two factors $SU(N)$ and $SU(N+M)$ run according to the beta functions: $\beta_{1/g_1^2}\sim -3M,\;\;\beta_{1/g_2^2}\sim 3M$. The two couplings run in opposite directions. Eventually $g_2$ will flow to strong coupling, requiring a Seiberg duality of the $SU(N+M)$ factor to be performed, reducing the gauge group down to $SU(N)\times SU(N-M)$. After that the two beta functions switch sign, and eventually $g_1$ diverges, requiring a Seiberg duality on the $SU(N)$ factor, further reducing the gauge group. This iterative cascade continues until the gauge group rank is smaller than $M$ and the theory confines. On the supergravity side this is manifested on the smooth capping off of the conical singularity in the deep IR (seen after solving for the back-reacted geometry). 

Since the original duality cascade of Klebanov and Strassler, there has been plenty of work aimed at generalizing the example. The two studies we will be most interested in are  \cite{Franco:2004jz} and \cite{Fiol:2002ah}. In \cite{Fiol:2002ah}, it was shown that the duality cascade can be reformulated in terms of the properties of the generalized Cartan matrix associated with the quiver gauge theory. Each Seiberg duality which is performed translates into a Weyl reflection in the associated root space.  It was further shown that for hyperbolic Cartan matrices, there generically exist ``duality walls." These walls are finite-energy scales in the UV which are accumulation points for the number of necessary Seiberg dualities. (In discussing duality cascades, one often tries to construct a candidate cascade flow by starting from the \emph{infrared} and moving into the \emph{ultraviolet}.) In this reformulation, it is easy to see that chaotic RG trajectories can occur, depending on the properties of the generalized Cartan matrix. This was first elucidated by \cite{Franco:2004jz}, where fractal characteristics of the approach to duality walls were shown. This connects chaotic behavior to hyperbolicity of the generalized Cartan matrix.

From a generalized Cartan matrix one can construct the associated Kac-Moody algebra. Hyperbolicity of the Cartan matrix corresponds to hyperbolicity of the underlying Kac-Moody algebra. The language of Weyl reflections in root spaces becomes equivalent to that of Weyl reflections in fundamental Weyl chambers of Kac-Moody algebras; this is precisely the language of the timelike BKL billiard we developed in Section \ref{coxeter}! Spacelike BKL billiards have a similar structure \cite{Henneaux:2007ej, Damour:2002et, Damour:2002fz, Damour:2000hv, Damour:2001sa}. Both in the case of a timelike BKL singularity and duality cascades, it is the \emph{hyperbolicity} of the underlying Kac-Moody algebra which implies chaos. This mathematical connection between our timelike BKL singularities and chaotic duality cascades provides some support for a renormalization group flow picture of our singularities in terms of a chaotic duality cascade.

Thus far, in the few dual supergravity flows of chaotic duality cascades that have been constructed, there seems to be no bulk signal of chaos. This is unsurprising, since individual bounces in the RG flow of the field theory are $\mathcal{O}(M/N)\sim \mathcal{O}(1/N)$ effects. The supergravity flow smooths over these bounces and thus the chaotic behavior is invisible in the classical bulk. The approach to a BKL singularity, however, is precisely the type of structure one would expect if the bulk geometry realized the flow of the boundary coupling constants. The Kasner exponents would mimic the coupling constants of the dual field theory, and they bounce around in the same way. In fact, \cite{Hanany:2003xh} exhibited a duality cascade in terms of a bouncing billiard, which upon every bounce switched the behavior of the coupling constants while always retaining the property that two were growing and one was shrinking. Such a field theory cascade is analogous to the BKL $D=4$ billiard, which in any Kasner epoch has two exponents growing while one is shrinking. 

What are we to make of the duality walls generically associated with hyperbolic quivers? These are accumulation points of Seiberg-duality transformations. The natural guess is to identify this with the BKL singularity in the bulk, since the bounces accumulate as one approaches the singularity. Unfortunately, to our knowledge duality walls have only been exhibited in the UV of field theories, so such a picture would not be appropriate for connecting our BKL regions to an asymptotically AdS spacetime. We stress, though, that the chaotic properties of BKL already call on a distinct notion of duality cascades of the boundary field theory which lead to visible bulk effects. One may hope for the existence of duality walls in the IR of such cascades. 

\subsection{Higher curvature corrections and the ultimate infrared}\label{highercurvcor}
To make sure we are in a controlled regime, we need  $g_s\ll 1$, $ r/\ell_s \gg 1$, and $r/\ell_{Pl} \gg 1$. In other words, perturbative string theory is valid and curvatures are small in string and Planck units. The BKL limit is naively in tension with this, since the simplest way to implement it is to zoom into the singularity, in which case higher curvature corrections would become important (see \cite{Damour:2005zb, Damour:2006ez} for work toward incorporating higher curvature corrections). However, by picking the length scale of inhomogeneities large compared to the Planck scale, and picking matter content with scales which lead to decoupling well before the singularity, we can open up parametric regions within which the BKL regime exists.

Given that we cannot approach arbitrarily close to the singularity, we have only commented on some intermediate phase of the RG flow. The ultimate fate of the RG flow is unknown to us and cannot be provided by a BKL-type analysis. It is possible that any resolution of the bulk singularity leads to a picture with  a chaotic RG flow that is transient. In other words, for finite energy decades the system looks chaotic but over a large enough energy scale the behavior disappears. Such behavior has been extensively studied in dynamical systems; a dissipative double pendulum would be a simple analogy for such transient chaos.

\section{Discussion}\label{conclusions}
Belinski, Khalatnikov, and Lifshitz wondered whether the singularity of FRW geometries is a generic feature of Einstein's equations or an artifact of the large symmetry group of that spacetime (rotations and translations). Similarly, we wondered whether the constructed supergravity flows which often seem to end in singular interiors are simply an artifact of the large symmetry group of the flows (often Lorentz invariance).

This paper was constructed with two goals in mind. First, purely as an exercise in general relativity, we constructed the timelike versions of the BKL singularities and analyzed their properties. On a more speculative front, we analyzed the implications of such singularities if they were to appear in the infrared of some AdS/CFT renormalization group flows. A key point to keep in mind when thinking about embeddings in AdS/CFT is the \emph{genericity} of the BKL picture. It is often possible to engineer bizarre scenarios, but genericity arguments are rare and powerful. 

The generalized radial Kasner geometry \eqref{kasnergenr} was the starting point in studying the BKL-like properties of the singularity. We found, as in the usual BKL case, that there are two possibilities: the evolution either becomes exactly ultralocal, which is not the generic scenario, or the situation becomes approximately ultralocal, with oscillatory bounces perturbing the otherwise exact ultralocality. This latter situation is generic, in the sense that the constructed geometries are solutions of nonvanishing measure in the full space of solutions. The one key difference between the spacelike and timelike singularities is that the appearance of chaos does not simply extend to higher dimensions.

The rigorous construction of BKL singularities is extremely difficult, so we did not attempt to realize any sort of flow from an asymptotically AdS geometry to a timelike BKL singularity. Nevertheless, we analyzed the implications of a putative flow containing a timelike BKL singularity. The immediate objection one might raise, which can apply to this work at large, is that these naked timelike singularities are pathological. Often one appeals to cosmic censorship to dismiss such geometries (although we were not able to provide evidence for such a situation, it is still possible that such timelike BKL singularities may exist behind horizons). However, whether such singularities are pathological is not clear to us in the context of AdS/CFT. We made an attempt to rule out the simple version of these geometries \eqref{kasnerr} in Section \ref{causality} by arguing that they would lead to acausal behavior in the dual field theory. This criterion was effective in ruling out the majority of these geometries, leading to the constraint $p_t > p_i$ for all spatial directions $i$, but it was unable to rule out all of the geometries. It is unclear what sorts of generalized radial Kasner geometries \eqref{kasnergenr} the causality criterion would rule out without explicit examples. It is possible that any sort of chaotic scenario can be ruled out, since such scenarios naturally include epochs with $p_t > p_i$. In a separate analysis, we showed that there do not seem to exist any extremal surface barriers which would prevent the boundary field theory from accessing the physics of the singularity. We thus speculated that the natural boundary interpretation of a flow ending in \eqref{kasnergenr} is in terms of an ultralocal renormalization group flow (in the case of exact ultralocality in the bulk IR), or an ultralocal and chaotic renormalization group flow (in the case of chaotic oscillatory bounces in the bulk IR). This speculation reached a fever pitch when we connected our chaotic singularities to the chaotic duality cascades of \cite{Franco:2004jz}. Interestingly, such timelike BKL singularities cannot exist for pure gravity in asymptotically AdS$_3$, possibly indicating that Lorentz-symmetry-breaking renormalization group flows of two-dimensional CFTs are special.

There are two key directions left open by our analysis. The first is to analyze the effects of matter which does not decouple near the singularity, with an aim toward classifying the chaotic properties of supergravity theories. The second is to do a more careful analysis of higher dimensions to decisively conclude the general dynamics of the symmetry wells and gravity wells. In particular, resolving the gravity wells in a representative example will help deduce the fate of the dynamics and possibly even lead to a restoration of chaos. 

There are many ways the arguments in this paper can go wrong. The initial data problem on a timelike slice may be ill-posed in certain instances. The singularities may be impossible to cloak with a horizon, which would be a problem if cosmic censorship were a principle of nature or AdS/CFT. Top-down embeddings may not exist. Numerical simulations may not support our picture. And so on. However, if the suggestive arguments of this paper uphold rigorous scrutiny, then within AdS/CFT they may hint at a beautiful new picture of generic renormalization group flows of holographic CFTs.

\section*{Acknowledgements}
We would like to thank Dionysios Anninos, Thibault Damour, Sean Hartnoll, Gary Horowitz, Shamit Kachru, Axel Kleinschmidt, Lampros Lamprou, Hermann Nicolai, and Gonzalo Torroba for useful discussions.  This work was supported by NSF Grants PHY-0756174 and PHY13-16748.

\appendix
\section{Spacelike BKL summary}
In this appendix we provide a brief summary of the heuristic arguments that led BKL to conjecture their picture of a spacelike singularity. The three upcoming sections summarize the arguments of \cite{Lifshitz:1963ps}, \cite{Belinsky:1970ew}, and \cite{Belinsky:1982pk}, respectively. In the main text we will begin with the same sorts of arguments for our timelike singularity before switching to the billiard picture of \cite{Damour:2002et}. This latter picture is more robust and allows one to easily consider arbitrary dimension and include the effects of dilatons and $p$-forms. It is also the picture which makes clear that timelike BKL singularities are generically less chaotic than spacelike BKL singularities. 

We will consider Einstein's equations in vacuum and will comment on the role of matter in \ref{appmatter}.

\subsection{No oscillations}\label{appkl1}
We begin by considering the less generic case of no oscillation, first considered in \cite{Lifshitz:1963ps}. We work in a synchronous reference system and expand the metric about the singularity as 
\be
ds^2=-dt^2+(a^2l_i l_j+b^2m_i m_j+c^2n_i n_j)dx^i dx^j=-dt^2+h_{ij}dx^i dx^j\,,\label{kasnert}
\ee
where $a$, $b$, $c$, $l_i$, $m_i$, and $n_i$ all depend on $t$ and $x_i$. The metric is seen to be Lorentzian by computing the determinant in the basis given by $\{l_i,m_j,n_k\}$, which gives det $g_{\mu\nu}=-(\epsilon^{ijk}l_i m_j n_k\, abc)^2$. Notice it is in general not possible to align the $x_i$ with the directions $l_i$, $m_i$, and $n_i$.

Defining $\kappa_i^j=\partial h_i^j/\partial t$, we can write Einstein's equations in this frame as 
\begin{align}
R_t^t&=-\frac{1}{2}\frac{\p \kappa_i^i}{\p t} -\frac{1}{4}\kappa_i^j\kappa_j^i=0\,, \label{firsteq}\\
R_i^j&=-\frac{1}{2\sqrt{|g|}}\frac{\p}{\p t}(\sqrt{|g|}\kappa_i^j)-^{(3)}R_i^j=0\,,\label{secondeq}\\
R_i^t&=\frac{1}{2}\left(\nabla_j\kappa_i^j-\nabla_i\kappa_j^j\right)=0\,.\label{thirdeq}
\end{align}
The first equation only has derivatives with respect to $t$. The same goes for the second equation if we can ignore $^{(3)}R_i^j$, the three-dimensional Ricci tensor built out of $h_{ij}$. Keeping it will be the subject of the next subsection. 

 Ignoring $^{(3)}R_i^j$, \eqref{firsteq} - \eqref{secondeq} become 
\begin{align}
R_t^t=\frac{a''}{a}+\f{b''}{b}+\f{c''}{c}&=0\,,\\
R_l^l=-\f{(a'bc)'}{abc}=0\,, \qquad R_m^m=-\f{(ab'c)'}{abc}&=0\,, \qquad R_n^n=-\f{(abc')'}{abc}=0\,,
\end{align}
where we have defined projections of the tensor $R_{\mu\nu}$ as $R_{mn}=R_{\mu\nu}m^\mu n^\nu$. These equations have as solution the anisotropic Kasner geometry \eqref{kasner} - \eqref{exps}. The final equation \eqref{thirdeq} can be reduced to three $t$-independent constraints on the exponents and vectors \cite{Lifshitz:1963ps}. As we will see in the next subsection, ignoring $^{(3)}R_i^j$ requires a single constraint. This additional constraint makes the parameterization a set of measure zero in the full phase space; keeping $^{(3)}R_i^j$ nonvanishing will instead allow the ansatz to cover an open set.\footnote{In fact, an interesting historical anecdote: the authors of \cite{Lifshitz:1963ps} originally ignored this possibility and concluded that their inability to parameterize a generic singularity meant that singularities were not a generic result of Einstein's equations! It was only after Penrose laid down the law with his singularity theorem \cite{Penrose:1964wq} that the authors revisited their assumptions in \cite{Belinsky:1970ew}.} 

The equations \eqref{firsteq}-\eqref{thirdeq} can also be recast in the Hamiltonian formalism. In this framework, the evolution equations are analogous to \eqref{firsteq} and \eqref{secondeq}, while the constraint equations are analogous to \eqref{thirdeq}. Ultralocality is then the statement that the spatial derivative terms in the \emph{evolution} equations are irrelevant asymptotically close to the singularity. This means that the evolution equations break up pointwise, and it allows us to zoom in to different points and track their evolution independently from nearby points. In other words, the set of partial differential equations governing the evolution of the three-metric become ordinary differential equations. Even in the case where the term $^{(3)}R_i^j$ has spatial derivatives which can compete, it is possible to treat the evolution pointwise by zooming in to different points or ``patches" of spacetime. We will refer to that situation as ``approximately ultralocal" and address it in the next subsection.

If we imagine the spacelike singularity exists in the future, then nearby points will causally decouple as one approaches the singularity due to the existence of horizons. This is an intuitive explanation often given to explain the ultralocal behavior of BKL. 

\subsection{Oscillations}\label{appbkl2}
The oscillatory case, first considered by BKL in \cite{Belinsky:1970ew}, generalizes the analysis of the previous section. To begin, we consider a homogeneous version of \eqref{kasnert} with specific scale factors:
\be\label{eq: time_Kasner}
ds^2=-dt^2+(t^{2p_1(t)}l_{i}l_{j}+t^{2p_2(t)}m_{i}m_{j}+t^{2p_3(t)}n_{i}n_{j})dx^{i}dx^{j}\,.
\ee
We will call this quasi-Kasner due to the time dependence of the Kasner exponents. This ansatz contains four gauge-fixed free parameters: three spacelike vectors each with three components, and the exponents $\{p_1(t), p_2(t), p_3(t)\}$, which are determined by a single parameter $u(t)$ through $p_1=\frac{-u}{1+u+u^2},\; p_2=\frac{1+u}{1+u+u^2},\; p_3=\frac{u+u^2}{1+u+u^2}$. In total we have ten parameters. However, we also have three of Einstein's equations $G_{ti}=0$ acting as constraints, as well as diffeomorphisms that allow redefining the three spatial coordinates, giving $10-3-3=4$ free parameters. This equals the number of free parameters in a generic homogeneous solution to the Einstein vacuum equations. Now consider $\lambda\equiv l^i \epsilon_{ijk}\partial_{x_j}l^k$, where $\vec{l}$ is the expanding direction. If $\lambda=0$, then we have $u(t)=\text{const}$, whereas we will have an oscillatory $u(t)$ for generic non-zero $\lambda$ (we will illustrate the oscillatory behavior below). The former case corresponds to imposing an additional constraint and therefore is not generic. We conclude that the generic behavior for a homogeneous approach to the singularity takes the quasi-Kasner form of (\ref{eq: time_Kasner}), but with oscillatory exponents.   

Due to the ultralocality near the singularity, we can view a generic evolution toward the singularity as consisting of different local ``patches" of quasi-Kasner spacetime. Therefore, in the BKL scenario, the four free parameters are uplifted in a generic approach toward the space-like singularity to four free functions of the spatial coordinates. This is the same number of degrees of freedom as needed to define a generic initial condition for Einstein's equations \cite{wald}. It is in this sense that BKL is considered a generic description of the singularity: it captures a set of non-zero measure of all possible spacetimes that produce a spacelike singularity.

Now let us analyze more closely the oscillatory dynamics by zooming in to one of the patches:
\be
ds^2=-dt^2+(a(t)^2l_{i}l_{j}+b(t)^2m_{i}m_{j}+c(t)^2n_{i}n_{j})dx^{i}dx^{j}\,.
\ee 
Einstein's equations take the following form:
\bea\label{eqnsosc}
\frac{(a'bc)'}{abc}=-\frac{1}{2a^2b^2c^2}[\lambda^2 a^4-(\mu b^2-\nu c^2)^2]\,,\nonumber\\
\frac{(ab'c)'}{abc}=-\frac{1}{2a^2b^2c^2}[\mu^2 b^4-(\lambda a^2-\nu c^2)^2]\,,\nonumber\\
\frac{(abc')'}{abc}=-\frac{1}{2a^2b^2c^2}[\nu^2 c^4-(\lambda a^2-\mu b^2)^2]\,,\nonumber\\
\frac{a''}{a}+\frac{b''}{b}+\frac{c''}{c}=0\,,
\eea
where $\lambda= l^i \epsilon_{ijk}\partial_{x_j}l^k$, $\mu= m^i \epsilon_{ijk}\partial_{x_j}m^k,$ and $\nu= n^i \epsilon_{ijk}\partial_{x_j}n^k$  can be functions of the spatial coordinates. We can approximate each such patch by some homogeneous (possibly anisotropic) geometry, i.e. one of the Bianchi types. Then ($\vec{l}, \vec{m}, \vec{n}$) can be identified with the Killing vectors generating the local isometry that characterizes the homogeneous geometry. And ($\lambda,\mu,\nu$) are related to the structure constants $C^r_{pq}$ of the Bianchi Type, defined by $[\vec{p},\vec{q}]=C^r_{pq}\vec{r}$:
\bea
\lambda = \epsilon^{lmn}C^l_{mn},\quad \mu=\epsilon^{mln}C^m_{ln},\quad \nu=\epsilon^{nml}C^n_{ml}\,.
\eea
For example, the case of $\lambda=\mu=\nu=0$ corresponds to flat space, while $\lambda=\mu=\nu=1$ corresponds to Bianchi Type VII.

In the previous subsection, we solved \eqref{eqnsosc} in the limit where $\lambda=\mu=\nu=0$. This is precisely the case where we set $^{(3)}R_i^j=0$ in \eqref{secondeq}, as advertised. We will now consider the more general case of $\lambda, \mu,\nu\neq 0$. In this case, a pure Kasner solution of $a(t),b(t),c(t)$ will cease to be valid at some point, since one of the scale functions, say $a(t)$, will be diverging instead of vanishing toward the singularity. This corresponds to the direction along $\vec{l}$ expanding instead of contracting. This will give a big contribution to the right-hand-side of \eqref{eqnsosc}, and we can capture its effect by keeping the dominant $a^4$ term and tracking how it modifies the Kasner solution. 
 
To see this, define $a=e^{\alpha},\,b=e^{\beta},\,c=e^{\gamma}$, and redefine $d\tau=dt/(abc)$. We begin with a Kasner regime with $\alpha= p_1\tau,\, \beta= p_2\tau$, and $\gamma= p_3\tau$. This gives  $d\tau= d (\ln \,t)$ since $abc=t$ in the initial Kasner epoch. Keeping the $a^4$ term on the right-hand-side of \eqref{eqnsosc} leads to the equations of motion 
\be\label{eq:bouncing}
2\,\f{d^2\alpha}{d\tau^2}=-\lambda a^4,\qquad2\,\f{d^2\beta}{d\tau^2}=\lambda a^4,\qquad2\,\f{d^2\gamma}{d\tau^2}=\lambda a^4\,.
\ee  
 The equations can be made more clear in the following form: 
\be
\f{d^2\alpha}{d\tau^2}=-\frac{e^{4\alpha}}{2}\,,\qquad\f{d^2\alpha}{d\tau^2}+\f{d^2\beta}{d\tau^2}=0\,,\qquad \f{d^2\alpha}{d\tau^2}+\f{d^2\gamma}{d\tau^2}=0\,.
\ee
These equations can be interpreted as the dynamics of three particles of unit mass at positions $\alpha, \beta, \gamma$. Now notice that the initial Kasner exponents are simply the momenta of the particles $d\alpha/d\tau = p_1,\, d\beta/d\tau= p_2,\, d\gamma/d\tau= p_3$.

The first equation shows that the first particle is colliding against an exponential wall, while the second two equations show that the total momentum of the pairs ($\alpha,\beta$) and ($\alpha,\gamma$) is conserved. This setup represents our system in a single Kasner epoch, and we would like to track it through a bounce due to the exponential potential. The relationship between $\tau$ and $t$ will no longer be $\tau=\ln t$ after a bounce. This means that while the incoming momenta were exactly equal to the Kasner exponents, the outgoing momenta will have to be rescaled to give the Kasner exponents. In practice, this rescaling is completely determined by ensuring that the Kasner exponents sum to $1$. 

The solution to this system of equations is as follows: the particle at position $\alpha$ will go through an elastic collision against the potential and bounce off with opposite momentum $p'_1=-p_1$. Momentum conservation then fixes the outgoing momenta for $\beta$ and $\gamma$: $p'_2=p_2+2p_1,$ $p'_3=p_3+2p_1$. In fact, the system can be solved analytically:
\bea\label{eq:bouncing-solution}
\alpha(\tau)&=&\frac{1}{2}\ln \left[2 p_1 \cosh \left(2 p_1(\tau-\tau_0)\right)^{-1}\right]\,,\nonumber\\
\beta(\tau)&=& (p_1+p_2)\tau-\alpha(\tau) + \text{const}\,,\nonumber\\
\gamma(\tau)&=& (p_1+p_3)\tau -\alpha(\tau)+ \text{const}\,,
\eea
where $\tau_0$ is the point of deflection, determined by the initial conditions. This solution has the correct asymptotic behavior away from $\tau_0$. We conclude that a generic non-vanishing right-hand-side in \eqref{eqnsosc} will cause the evolution to switch from one set of Kasner exponents to another set:
\be
(p_1,p_2,p_3)\longrightarrow (\tilde{p}_1,\tilde{p}_2,\tilde{p}_3)=\left(\frac{-p_1}{1+2p_1}, \frac{p_2+2p_1}{1+2p_1},\frac{p_3+2p_1}{1+2p_1}\right)\,.
\ee
As argued earlier, the exponents require a universal normalization due to the relation $d\tau=dt/(abc)$. To really put the system into the Kasner form \eqref{eq: time_Kasner} after a bounce, we have also rescaled constant prefactors into the vectors $\vec{l}$, $\vec{m}$, $\vec{n}$. 

The transition is such that the expanding direction switches to contracting, and one of the contracting directions, say $b(t)$, begins to expand instead. The new Kasner solution will remain valid for a while until the right-hand-side becomes relevant due to a dominant $b^4$ term, and the system will make yet another transition. This oscillatory process repeats itself in the BKL scenario and leads to chaotic behavior. 

\begin{figure}
\center{\includegraphics[scale=.3]{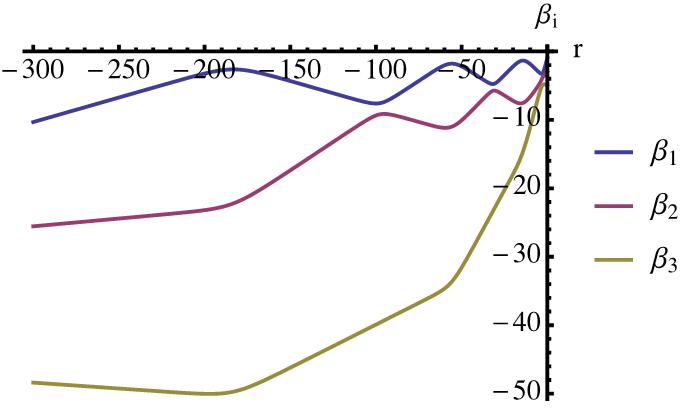}\qquad \includegraphics[scale=.3]{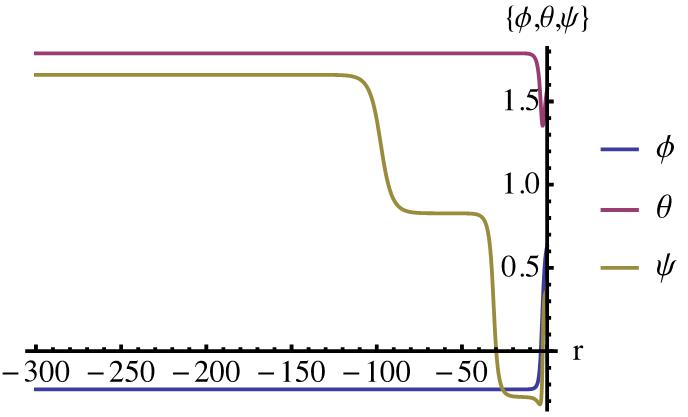}}
\caption{Left: dynamics of the Iwasawa exponents $\beta_i$, and the evolution toward the singularity is in the negative $t$ direction. We see that the ordering of the exponents is maintained, a necessary result of the Iwasawa frame. Right: dynamics of the angular variables coming from the off-diagonal modes.}\label{spacelike_curv_and_sym}
\end{figure}

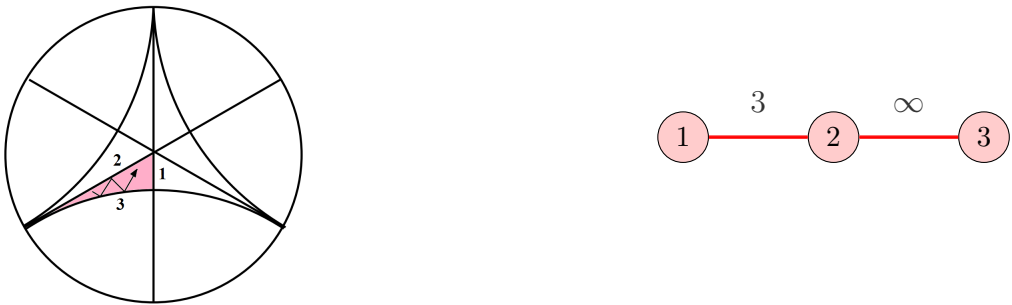
\begin{figure}

\begin{minipage}[h]{0.5\linewidth}\hspace{1.2cm}
\includegraphics[scale=.2]{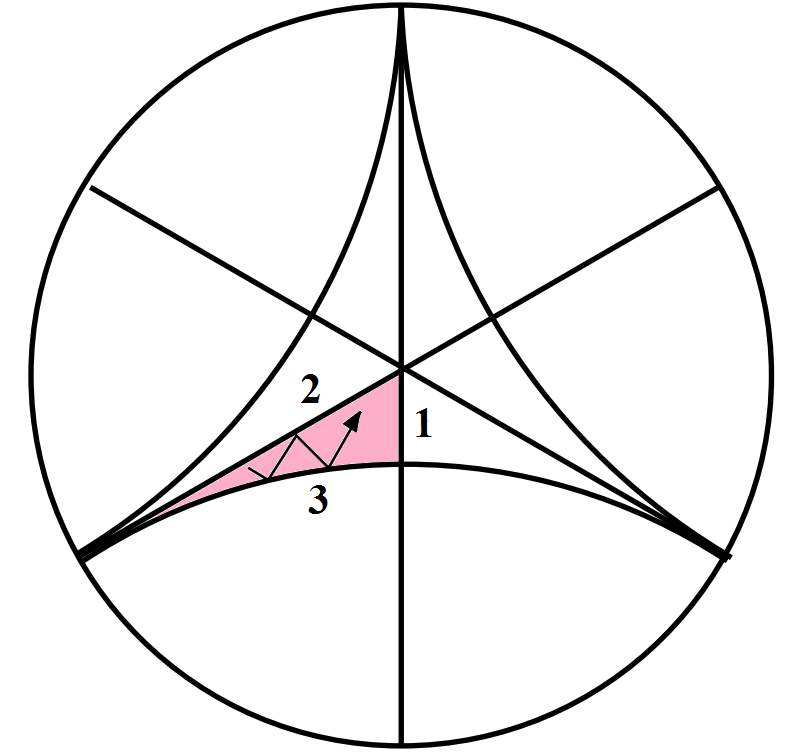}
\label{fig:figure1}
\end{minipage}
\hspace{0.5cm}
\begin{minipage}[b]{0.4\linewidth}
\centering
\begin{tikzpicture}\vspace{-3cm}
 \tikzset{LabelStyle/.style =   {                                text           = black}}

\tikzstyle{line} = [draw=red, -latex']
  \node[fill=red!20,draw,circle] (n1) at (0,0) {1};
  \node[fill=red!20,draw,circle]  (n2) at (2,0)  {2};
  \node[fill=red!20,draw,circle]  (n3) at (4,0)  {3};

 \draw[-, >=latex, color=red!80,double=red] (n1) to node[LabelStyle,above=.2cm]{{\large $3$}} (n2) ;
 \draw[-, >=latex, color=red!80,double=red](n2) to node[LabelStyle,above=.2cm]{{\large$\infty$}} (n3);
\end{tikzpicture}
\end{minipage}
\caption{This figure, taken from \cite{Henneaux:2007ej}, shows one of the confining regions of the pure four-dimensional gravity billiard. In the language of Section \ref{billiards}, the straight lines are symmetry walls and the curved lines are gravity walls. The Coxeter diagram representing the symmetry of reflections in this billiard is given on the right. The numbered nodes correspond to the reflection walls in the left-hand billiard.}\label{domains}
\end{figure}

In the billiard dynamics picture of Section \ref{billiards}, where one tracks the Iwasawa exponents, we have plotted the dynamics of the exponents in Figure \ref{spacelike_curv_and_sym}. In the same billiard dynamics picture, we have displayed the confining regions of the billiard stadium in Figure \ref{domains}.

\subsection{Non-diagonal spatial metrics}\label{appbkl3}
In the analysis of \ref{appbkl2}, it was assumed that the spatial metric could be diagonalized for all time. In other words, for a homogeneous metric of the form
\be
ds^2=-dt^2+\gamma_{ab}(t) (e_i^{(a)} dx^i)(e_j^{(b)}dx^j)\,,
\ee
where the vectors $e_i^{(1)}=l_i$, $e_i^{(2)}=m_i$, and $e_i^{(3)}=n_i$ are functions of the spatial coordinates, it was assumed that $\gamma_{ab}$ was a diagonal matrix of scale factors. Although this is a consistent truncation in the degrees of freedom, since the off-diagonal components of the Ricci tensor will remain vanishing, it is an unnecessary restriction.

In \cite{Belinsky:1982pk}, this restriction was lifted and more general properties of the chaotic approach to the singularity were explored. We will not explore their analysis since this general case is captured more elegantly by the billiard dynamics of \cite{Damour:2002et}, which we employ in the main text. The off-diagonal degrees of freedom map into dynamics of some angular variables, which we have simulated in Figure \ref{spacelike_curv_and_sym}.

\subsection{The effects of matter}\label{appmatter}
So far we have worked with the Einstein equations in vacuum. The addition of matter is a complicated issue whose story we do not aim to develop but will summarize. The original analysis in \cite{Lifshitz:1963ps} considered a relativistic fluid and showed that their contribution to \eqref{firsteq} - \eqref{secondeq} was subleading in $1/t$, while the contribution to \eqref{thirdeq} was merely to change the constraints on the exponents and vectors by including the free functions having to do with the matter. The result was still one arbitrary function short of the generic case with oscillation. Since this analysis, many different types of matter have been considered, in various dimensions, with the primary aim being the determination of whether or not the approach to the singularity is oscillatory. In $D\geq 11$ dimensions, the non-oscillating Kasner solutions are no longer one free function short of the general solution  \cite{Demaret:1986su}. In other words, even if oscillatory solutions exist, the non-oscillatory solutions are just as generic, in stark contrast with the lower-dimensional case. For $D<11$ dimensions, the BKL story seems to be robust except for a few exceptional cases. One of those exceptional cases involves a dilaton, which by itself stops the oscillatory nature of the singularity, although this can be countered by the effects of additional matter, e.g. $p$-form gauge fields \cite{Belinski:1973zz, Damour:2002tc}.

\section{Symmetry well artifacts in Bianchi IX}\label{symwells}
As stated in Section \ref{billiardsdetail}, the naive analytic continuation of the symmetry walls from the case of the spacelike BKL billiard does not work to give the correct symmetry walls in the case of a timelike BKL billiard. In $3+1$ dimensions, where symmetry walls seem to reflect the symmetries of the problem for the spacelike BKL billiard, we expect for two of the three walls to become transparent. These transparencies correspond to the two pairs of coordinates involving time which are not symmetric. The other wall involves a pair of spatial coordinates, which is symmetric both in the spacelike BKL case and our timelike BKL case.  The naive continuation shows that two of the walls become wells, with the rest of the walls untouched. So it gets most of the picture right. However, the approximation made in the billiard picture break down for a timelike singularity, so we need to revisit the equations of motion and solve for the exact outcome of these purported wells. We will do this in a Bianchi IX geometry, which will verify our physical expectations. Recall that Bianchi IX is general in the sense that it has the most number of nonvanishing structure constants possible. As in \cite{RyanJr1972301}, we will employ an orthogonal diagonalization of the three-metric $\gamma$ instead of the Iwasawa decomposition employed in Section \ref{billiards}, although the two are simply related as discussed in \cite{Damour:2002et}.

The metric takes the form
\be
ds^2 = |\gamma| dr^2+\gamma_{ij}(r) \omega^i\omega^j \,,
\ee 
where $\gamma_{ij}(r)$ is a Lorentzian signature three-metric. It is diagonalizable by orthogonal matrices:
\bea
\gamma_{ij}(r)&=& R(r)^{-1} \Gamma(r)  R(r)\nonumber\\
R(r)&=&R_{\psi}(r)R_{\theta}(r)R_{\phi}(r)
\eea

$R_\psi(r)=
\left( \begin{array}{ccc}
\cos{\psi(r)} & \sin{\psi(r)} & 0 \\
-\sin{\psi(r)} & \cos{\psi(r)} & 0 \\
0 & 0 & 1 \end{array} \right),\;\; R_\theta(r)=
\left( \begin{array}{ccc}
1 & 0 & 0 \\
0 & \cos{\theta(r)} & \sin{\theta(r)} \\
0 & -\sin{\theta(r)} & \cos{\theta(r)} \end{array} \right)$

$R_\phi(r)=
\left( \begin{array}{ccc}
\cos{\phi(r)} & \sin{\phi(r)} & 0 \\
-\sin{\phi(r)} & \cos{\phi(r)} & 0 \\
0 & 0 & 1 \end{array} \right),\;\;\Gamma(r)=
\left( \begin{array}{ccc}
\Gamma_1(r) & 0 & 0 \\
0 & \Gamma_2(r) & 0 \\
0 & 0 & -\Gamma_3(r) \end{array} \right)\\$
\\
The equations of motion take the form
\bea
&&-\frac{1}{|\gamma|}\kappa^{b'}_a(r)+P^b_a(r)=0\,,\nonumber\\
&&-\frac{1}{4|\gamma|}\left(\kappa^b_a\kappa^a_b-\frac{(\gamma')^2}{\gamma^2}\right)-P^a_a=0,\qquad \kappa^a_b \equiv\gamma^{bc}\gamma'_{ca}\,,
\eea
It satisfies the constraint $\kappa^a_b(r)'C^b_{ac}=0$, implying $\kappa^a_b(r)C^b_{ac}$ is a constant vector, which we take to be $(0,0,c)$\footnote{Strictly speaking, non-zero $c$ would violates the momentum constraint of the system in a homogeneous vacuum model, here we assume some matter/inhomogeneity is feeding in a non-zero $c$.}. The equations can then be simplified to
\bea
\phi' &=&-c\left\lbrace\frac{\Gamma_1\Gamma_3\cos^2{\psi}}{(\Gamma_1+\Gamma_3)^2}+\frac{\Gamma_2\Gamma_3\sin^2{\psi}}{(\Gamma_2+\Gamma_3)^2}\right\rbrace,\nonumber\\
\theta' &=&-c\sin{\theta}\cos{\psi}\sin{\psi}\left\lbrace\frac{\Gamma_2\Gamma_3}{(\Gamma_2+\Gamma_3)^2}-\frac{\Gamma_1\Gamma_3}{(\Gamma_1+\Gamma_3)^2}\right\rbrace,\nonumber\\
\psi' &=& c\,\cos\theta\left\lbrace\frac{\Gamma_1\Gamma_2}{(\Gamma_1-\Gamma_2)^2}+\frac{\Gamma_1\Gamma_3\cos^2{\psi}}{(\Gamma_1+\Gamma_3)^2}+\frac{\Gamma_2\Gamma_3\sin^2{\psi}}{(\Gamma_2+\Gamma_3)^2}\right\rbrace.\nonumber\\
\eea
The diagonal variable dynamics are described by
\bea
\left(\ln{\Gamma_1}\right)''&=& c^2\cos^2{\theta}\frac{\Gamma_1\Gamma_2(\Gamma_1+\Gamma_2)}{(\Gamma_1-\Gamma_2)^3}-c^2\sin^2{\theta}\cos^2{\psi}\frac{\Gamma_1\Gamma_3(\Gamma_1-\Gamma_3)}{(\Gamma_1+\Gamma_3)^3}+(\Gamma_2+\Gamma_3)^2-\Gamma_1^2\,,\nonumber\\
\left(\ln{\Gamma_2}\right)''&=& -c^2\cos^2{\theta}\frac{\Gamma_1\Gamma_2(\Gamma_1+\Gamma_2)}{(\Gamma_1-\Gamma_2)^3}-c^2\sin^2{\theta}\sin^2{\psi}\frac{\Gamma_2\Gamma_3(\Gamma_2-\Gamma_3)}{(\Gamma_2+\Gamma_3)^3}+(\Gamma_1+\Gamma_3)^2-\Gamma_2^2\nonumber\,,\\
\left(\ln{\Gamma_3}\right)''&=& c^2\sin^2{\theta}\cos^2{\psi}\frac{\Gamma_1\Gamma_3(\Gamma_1-\Gamma_3)}{(\Gamma_1+\Gamma_3)^3}+c^2\sin^2{\theta}\sin^2{\psi}\frac{\Gamma_2\Gamma_3(\Gamma_2-\Gamma_3)}{(\Gamma_2+\Gamma_3)^3}+(\Gamma_1-\Gamma_2)^2-\Gamma_3^2\nonumber\,.\label{diageqns}\\
\eea
These are the same equations as in \cite{RyanJr1972301}, except with $\Gamma_3\to -\Gamma_3$ which reflects the timelike nature of one of our coordinates. 

Now we can analyze the dynamics. In particular, we will focus on the dynamics from the symmetry potential, i.e. terms proportional to $c^2$. The gravity walls simply constrain $\Gamma_i < 1$. Assume we start in the BKL regime $\Gamma_1\gg\Gamma_2\gg \Gamma_3$, but then have the two spacelike directions $\Gamma_2$ and $\Gamma_1$ approaching each other. By this we mean $\Gamma_1/\Gamma_2 \sim \mathcal{O}(1)$ while $\Gamma_1-\Gamma_2 \sim \mathcal{O}(\Gamma_1)$. The symmetry potential takes the same form as in the spacelike BKL singularity. We will show that  this results in the two exponents $\beta_1=\ln{\Gamma_1}$ and  $\beta_2=\ln{\Gamma_2}$ colliding into one another and switching trajectories, which is precisely what happens in the case of a spacelike BKL singularity.

As $\Gamma_1 \to \Gamma_2$, we can keep the first term on the right-hand-side of the first two equations of \eqref{diageqns}. The dynamics is then given by
\be
\left(\beta_1+\beta_2\right)''\sim 0,\qquad \left(\beta_1-\beta_2\right)''\sim 2c^2\cos^2{\theta}\,e^{-(\beta_1-\beta_2)}.
\ee
The center-of-mass mode $\beta_1+\beta_2$ has conserved momentum, while the relative mode $\beta_{12}=\beta_1-\beta_2$ reflects off the exponential potential $V(\beta_{12})\sim e^{-\beta_{12}}$ and changes sign. The end result is that $\beta_1$ and $\beta_2$ switch momenta and resume each other's would-be trajectories. This is the usual picture of a spacelike symmetry wall. Notice that the exact form of the symmetry potential starts deviating from the exponential form (in fact becomes singular) in the limit $\beta_1\to \beta_2$, but since this regime is screened by the dynamics we do not need to worry about it.

Now we examine what happens when the timelike exponent $\beta_3=\ln{\Gamma_3}$ approaches $\beta_2$ (the case with $\beta_3, \beta_1$ is analogous). As before, we mean $\Gamma_2/\Gamma_3 \sim \mathcal{O}(1)$ and $\Gamma_2-\Gamma_3 \sim \mathcal{O}(\Gamma_2)$. Again keeping the leading terms, the dynamics is given by
\be
\left(\beta_2+\beta_3\right)''\sim 0,\qquad \left(\beta_2-\beta_3\right)''\sim -2c^2\sin^2{\theta}\sin^2{\psi}\,e^{-(\beta_2-\beta_3)}
\ee
The center-of-mass mode $\beta_2+\beta_3$ has conserved momentum, while the relative mode $\beta_{23} = \beta_2-\beta_3$ is attracted into an exponential well $V(\beta_{23})\sim -e^{-\beta_{23}}$. This is what we correctly deduced by the naive analytic continuation in Section \ref{billiardsdetail}. However, since now the regime $\beta_2\to\beta_3$ is not screened by the symmetry potential, we have to resolve the exponential potential into its exact form (ignoring irrelevant angular and $c^2$ prefactors):
\be 
V(\beta_{23})=-\frac{1}{2\cosh^2{\left(\frac{\beta_{23}}{2}\right)}}\,.
\ee
\begin{figure}
\center{\includegraphics[scale=.33]{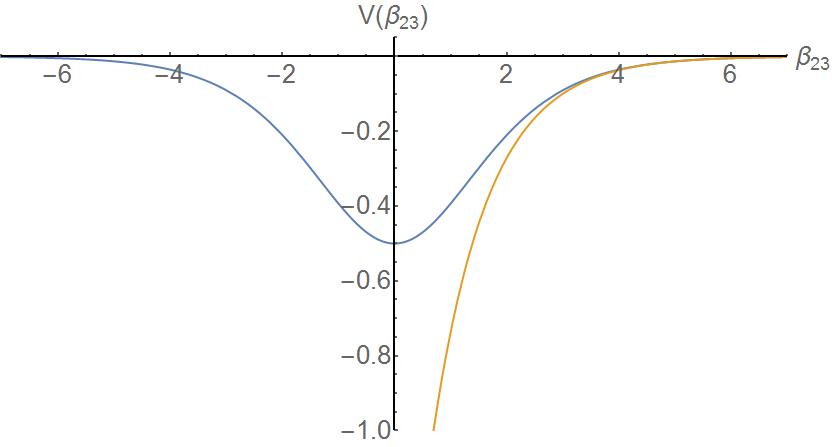}}
\caption{Exact potential (blue), compared with the asymptotic exponential potential (orange).}\label{exact_potential}
\end{figure}
Figure (\ref{exact_potential}) is a plot of the exact potential. As $\beta_{23}\to 0$, we have
\be 
V(\beta_{23})\sim -\frac{1}{2}+\frac{\beta_{23}^2}{8}\,.
\ee
The motion through the origin $\beta_3=\beta_2$ is therefore analogous to the way a harmonic oscillator passes through the origin, with vanishing force exerted. After that, the relative mode enters the other side of the regime $\beta_{23}<0$, and the asymptotic dynamics returns to free flight:
\be
\beta_{23}''\sim -e^{\beta_{23}}\to 0 \,.
\ee
The relative mode exits with the initial momentum, meaning $\beta_2, \beta_3$ pass through each other with the same momenta. What was originally a symmetry wall in the spacelike BKL singularity becomes transparent.  

\begin{figure}
\center{\includegraphics[scale=.32]{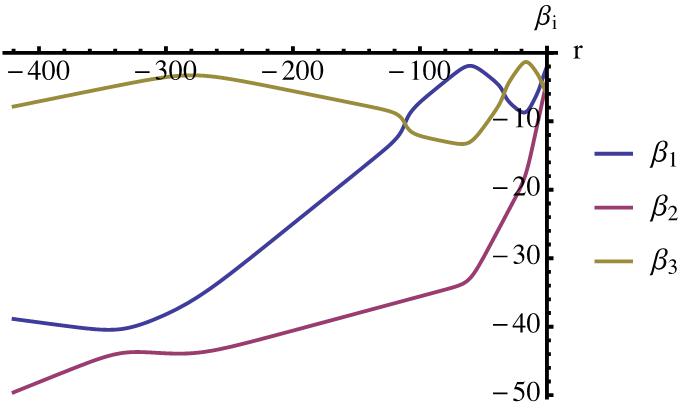}\qquad \includegraphics[scale=.32]{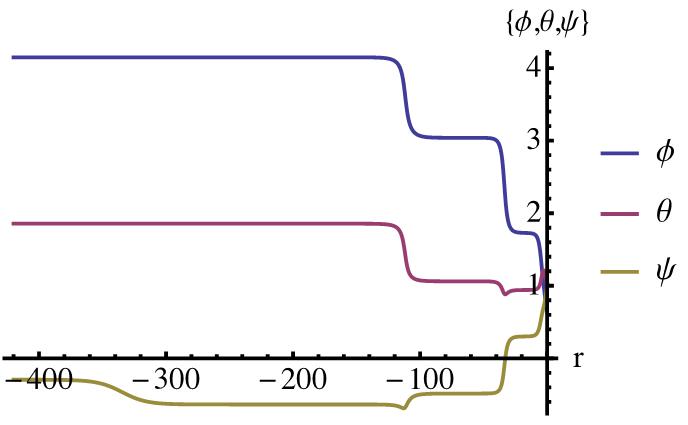}}
\caption{Left: dynamics of the diagonal exponents $\beta_i$ with $\beta_3$ the timelike direction. Notice how $\beta_1$ and $\beta_2$ collide in a way that switches momenta, while $\beta_3$ passes through $\beta_1$, consistent with the analysis. (It would also pass through $\beta_2$ but this is not illustrated.) Right: dynamics of the angular variables coming from the off-diagonal modes.}\label{gaussianplots}
\end{figure}

\begin{figure}
\center{\includegraphics[scale=.32]{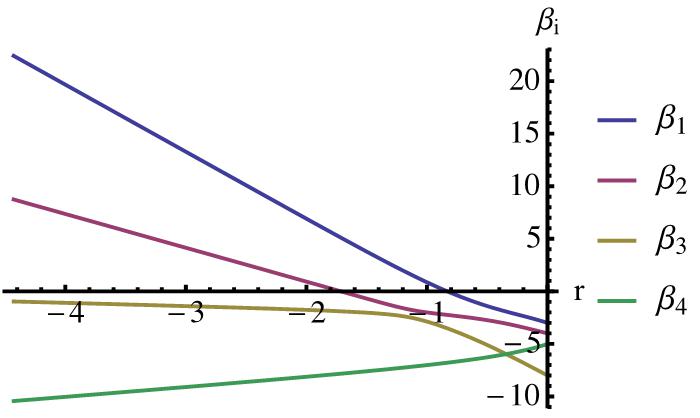}\qquad\includegraphics[scale=.32]{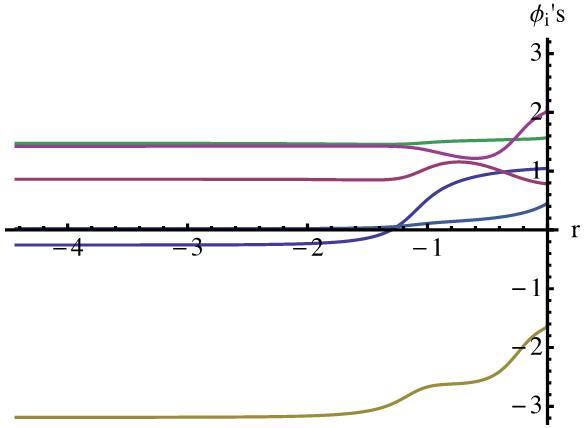}} 
\caption{Symmetry wall dynamics in 4+1 D systems. Left: dynamics of the diagonal exponents $\beta_i$ with $\beta_4$ the timelike direction. Right: dynamics of the angular variables $\phi_i$ coming from the off-diagonal modes.}\label{5D_sym_wall}
\end{figure}

To confirm this analysis, in Figure \ref{gaussianplots} we include a plot of the numerically generated dynamics for the diagonal variables as well as angular variables for some generic initial conditions. A picture of 4+1 dimensional dynamics including only the symmetry walls is shown in Figure \ref{5D_sym_wall}, which provides additional evidence that the picture regarding analytic continuation of symmetry wall dynamics is true in generic dimensions.  

\small{
\bibliography{BKLbib}
\bibliographystyle{apsrev4-1long}}

\end{document}